\documentclass[prl,twocolumn,superscriptaddress]{revtex4-2}

\usepackage{graphicx,xcolor}
\usepackage{float}
\usepackage[caption = false]{subfig}
\usepackage{amsmath}
\usepackage{amssymb}
\usepackage{mathtools}
\usepackage{hyperref}
\hypersetup{pdfencoding=auto,colorlinks=true,linkcolor=blue,citecolor=blue,urlcolor=blue}
\usepackage[normalem]{ulem}
\usepackage{soul}
\usepackage{dsfont}
\usepackage{textcomp}
\usepackage{tabstackengine}
\usepackage{colortbl}

\newcommand{\ket}[1]{|{#1}\rangle}
\newcommand{\bra}[1]{\langle{#1}|}

\newcommand{\Id}{\mathds{1}}
\DeclareMathOperator{\e}{e}

\newcommand\beq            {\begin{equation}}
\newcommand\eeq           {\end{equation}}
\newcommand\bwt         {\begin{widetext}}
\newcommand\ewt         {\end{widetext}}

\begin{document}

\title{Readout of parafermionic states by transport measurements}

\author{Ida E. Nielsen}
\email{ida.nielsen@nbi.ku.dk}
\affiliation{Center for Quantum Devices, Niels Bohr Institute, University of Copenhagen, DK--2100 Copenhagen, Denmark}
\affiliation{Niels Bohr International Academy, University of Copenhagen,
DK--2100 Copenhagen, Denmark}

\author{Karsten Flensberg}
\affiliation{Center for Quantum Devices, Niels Bohr Institute, University of Copenhagen, DK--2100 Copenhagen, Denmark}

\author{Reinhold Egger}
\affiliation{Institut f\"ur Theoretische Physik, Heinrich Heine Universit\"at, D-40225 D\"usseldorf, Germany}

\author{Michele Burrello}
\affiliation{Center for Quantum Devices, Niels Bohr Institute, University of Copenhagen, DK--2100 Copenhagen, Denmark}
\affiliation{Niels Bohr International Academy, University of Copenhagen,
DK--2100 Copenhagen, Denmark}

\begin{abstract}

Recent experiments have demonstrated the possibility of inducing superconducting pairing into counterpropagating fractional quantum Hall edge modes. This paves the way for the realization of localized parafermionic modes, non-Abelian anyons that share fractional charges in a nonlocal way. We show that, for a pair of isolated parafermions, this joint degree of freedom can be read by conductance measurements across standard metallic electrodes. We propose two complementary setups.
We investigate first the transport through a grounded superconductor hosting two interacting parafermions.
In the low-energy limit, its conductance peaks reveal their shared fractional charge yielding a three-state telegraph noise for weak quasiparticle poisoning.
We then examine the two-terminal electron conductance of a blockaded fractional topological superconductor, which displays a characteristic $e/3$ periodicity of its zero-bias peaks in the deep topological regime, thus signalling the presence of parafermionic modes.

\end{abstract}

\maketitle

\textit{Introduction.}---The most common experimental signatures of the existence of Majorana zero-energy modes (MZM) \cite{kitaev2001} are based on charge transport measurements of hybrid superconductor-semiconductor devices \cite{mourik2012,deng2016,nichele2017,kouwenhoven2018,marcus2018}.
These measurements can probe the existence of midgap excitations in topological superconductors, but do not provide a direct proof of their topological nature.
Furthermore, transport does not allow for a readout of the fermionic parity shared by a pair of MZM: when coupling MZM to an external lead, electrons with energies below the superconducting (SC) gap tunnel into the system in a process that continuously flips the Majorana fermionic parity. This enables the detection of subgap states at the price of losing the information encoded in their parity, which is the main degree of freedom adopted in the proposals for topological quantum computation based on MZM.
In this work, we show that this picture is fundamentally different when considering a fractionalized version of MZM, the so-called parafermionic zero-energy modes \cite{fendley2012} (parafermions for short). We provide a model for the tunnelling spectroscopy of two parafermionic SC devices and we show that transport measurements can be used to detect their shared degree of freedom.

\begin{figure}[h!]
\includegraphics[width=0.9\columnwidth]{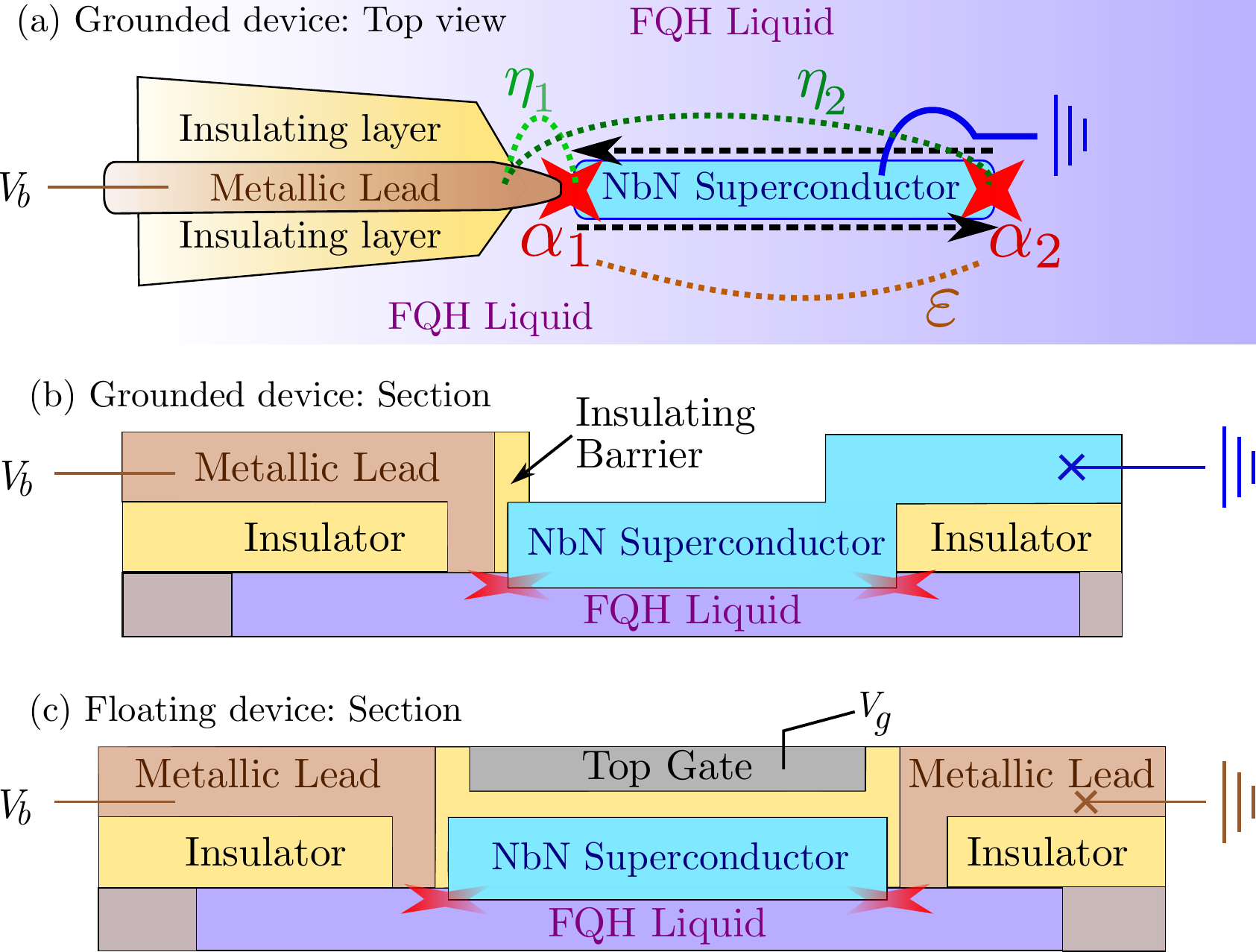}
\caption{Two-parafermion devices based on the SC pairing induced between chiral edge modes [dashed black arrows in (a)] by a NbN SC finger (light blue). (a) and (b) Top view and section of a grounded device. The NbN finger is in contact with the edges of a trench in a FQH liquid and defines two parafermions (red stars). A metallic lead (brown) is put in contact with the FQH liquid close to the left parafermion. The NbN finger is grounded through a SC bridge shown in (b). In (a) we schematically illustrate the tunnel couplings $\eta_{1,2}$ between lead and parafermions, and the parafermion overlap $\varepsilon$. (c) Coulomb blockaded device with a floating SC island. Its induced charge is controlled through the gate voltage $V_g$.} \label{fig:2pfsetup}
\end{figure}

Parafermions are localized topological excitations predicted to emerge in several heterostructures engineered by suitably coupling counterpropagating edge modes of fractional quantum Hall (FQH) states through an induced SC pairing \cite{stern2012,shtengel2012,vaezi2013,cheng2013,prophecy,barkeshli2014}. This is a demanding objective since the strong magnetic fields required by FQH systems typically suppress the coherence of the necessary SC elements. Indeed, only very recently the combination of these two main ingredients has been experimentally achieved in a graphene-based Hall device contacted with a SC NbN electrode \cite{kim2020} (see also Ref.~\cite{kim2017}). This system displayed evidence of crossed Andreev reflection of fractional quasiparticles, a key feature of parafermions hybridized with quantum Hall edge modes \cite{alicea2014}.
In the simplest case, based on the $\nu = 1/3$ Laughlin state, the resulting parafermionic topological superconductor is characterized by an emerging $\mathbb{Z}_6$ symmetry: a pair of parafermions causes a sixfold degeneracy of its ground states, which can be distinguished by their fractional charge $Q = q e/3$, with $q=0,\ldots,5$, such that $Q$ is defined modulo the Cooper pair charge \cite{stern2012,shtengel2012}.

Building on the experimental setup of Ref.~\cite{kim2020}, we consider devices in which two parafermions are in contact with external electrodes (Fig.~\ref{fig:2pfsetup}). The coupling between parafermions and metallic leads allows for electrons to tunnel in and out of the topological superconductor: the integer part of $Q$ is therefore not conserved, but the fractional charge $\tilde{Q} = q e/3 \mod e$ is a conserved and topologically protected quantity, which can store quantum information and be read through transport measurements.

In the following, we present two complementary parafermion systems (see Fig.~\ref{fig:2pfsetup}): First, we discuss a grounded device with interacting parafermions, whose energy splitting allows us to monitor their fractional charge $\tilde{Q}$ based on conductance measurements. Then, we examine a two-terminal blockaded setup with sizable charging energy; in the opposite limit of noninteracting parafermions, such system displays characteristic conductance signatures caused by these topological modes.

\textit{Modelling of the parafermions.}---We begin our analysis by considering a SC device hosting two isolated parafermions. Such a device can be fabricated based on the techniques adopted in Ref.~\cite{kim2020}. A NbN SC thin finger is deposited in a trench etched in the bulk of a FQH liquid (Fig.~\ref{fig:2pfsetup}). Two counterpropagating FQH edge modes appear on either side of the trench and are gapped by the proximity-induced superconductivity, allowed by the strong spin-orbit coupling of NbN. Differently from Refs.~\cite{alicea2014,kim2020}, we consider a device in which the superconductor is not coupled to any additional edge mode of the FQH liquid, such that the two resulting parafermions, $\alpha_1$ and $\alpha_2$, are isolated: given the bulk gap of the FQH liquid, they do not exchange fractional charges with the environment at low temperature.

The parafermionic operators $\alpha$ obey the following rules \cite{shtengel2012}:
\begin{equation}
\alpha_2 \alpha_1 = \e^{-i\pi/3}\alpha_1\alpha_2\,, \qquad
\alpha_2^\dag \alpha_1 = \e^{i\pi/3}\alpha_1\alpha_2^\dag\,, \label{pfcomm}
\end{equation}
with $\alpha_i^6=\Id$ and $\alpha_i^\dag = \alpha_i^{-1}$. The zero-energy parafermions $\alpha_1$ and $\alpha_2$ induce a sixfold degeneracy of the ground states of this device which can be characterized by a suitable $\mathbb{Z}_6$ parity \cite{cobanera2013}: $\e^{-i\pi/6}\alpha_2^\dag \alpha_1= \e^{-i\pi q/3}$. Here $q$ is a number operator with eigenvalues $0,\ldots,5$ counting fractional charges (mod $6$) in the segment of the counterpropagating edge modes coupled with the superconductor and represents a joint observable of the two parafermions.

For energies below the induced SC gap, a weak coupling $\varepsilon$ between $\alpha_1$ and $\alpha_2$ lifts the ground state degeneracy,
\begin{equation}
H_{\rm pf} = -\varepsilon \e^{-i\left(\pi/6+\phi\right)} \alpha^\dag_2 \alpha_1 + {\rm H.c.} = -2\varepsilon\cos\left(\pi q/3+\phi\right).
\label{eq:H_pf}
\end{equation}
$\varepsilon$ indicates the overlap of the two modes and $\phi \propto \mu L$ is a phase that depends on the chemical potential $\mu$ of the chiral edge modes and the SC finger length $L$ \cite{burnell2016,schmidt2019}.

\textit{The grounded N-SC device.}---We couple this device with an external metallic lead [see Figs.~\ref{fig:2pfsetup}(a) and \ref{fig:2pfsetup}(b)] and introduce the ladder operators $l$ and $l^\dag$ associated with electrons at the tunnel contact (the strong magnetic field polarizes the parafermions, therefore we consider only one spin species). The coupling between lead and parafermions causes both the coherent tunnelling process of one electron moving into the superconductor, and the formation of a Cooper pair from an electron in the lead and an electron extracted from the FQH edges. The operators $\alpha_i^3 = (\alpha_i^\dag)^3 \equiv \gamma_i$ anticommute and define two localized MZM. Therefore, in analogy with similar Majorana setups \cite{beenakker2008,fu2010,flensberg2010,egger2011,fidkowski2012}, we model the coupling as
\begin{equation} \label{coupling}
H_c = i \sum_{j=1,2 }\eta_j \gamma_j \left(\e^{i \chi_j} l + \e^{-i \chi_j} l^\dag\right)\,,
\end{equation}
where $\eta_{1/2}\e^{i\chi_{1/2}}$ is the tunnelling amplitude between the lead and the left/right parafermion [Fig.~\ref{fig:2pfsetup}(a)] with $\{\eta_j,\chi_j\}$ real and typically $\eta_1 \gg \eta_2$ (more general couplings and their renormalization group relevance are considered in the Supplemental Material (SM) \cite{suppl}). $H_c$ changes the charge $Q$ by $\pm e$ and the Majorana operators $\gamma_j$ allow us to rewrite $H_{\rm pf}$ by separating a fermionic and a fractional charge. The number $q$ of fractional charges (modulo $6$) can indeed be expressed in terms of the parity $p=i\gamma_2\gamma_1=\e^{i\pi q}$ and the fractional charge $e\tilde{q}/3=eq/3  \mod e$ (i.e., $\tilde{q}=0,1,2$), such that each eigenstate $\ket{q}$ is recast in the form $\ket{p,\tilde{q}}$ (specifically, $\e^{i\pi q/3} = p\e^{i\pi 4\tilde{q}/3}$ \cite{suppl}). The parafermion coupling becomes $H_{\rm pf} = -i2\varepsilon \gamma_2\gamma_1 \cos(4\pi\tilde{q}/3+\phi)$, and is thus quadratic in the Majorana operators, while the charge $e\tilde{q}/3$ is a conserved quantity commuting with $H_c$. 

As a result, the conductance of the two-terminal device in Figs.~\ref{fig:2pfsetup}(a) and \ref{fig:2pfsetup}(b), between the normal lead and the grounded SC electrode, is derived through the corresponding two-Majorana setup \cite{flensberg2010} and is estimated by applying the Weidenm\"uller formula \cite{aleiner2002} for each value of the conserved fractional charge $e\tilde{q}/3$ separately \cite{suppl}. In the low-energy approximation defined by $H=H_{\rm pf} + H_c$, valid when the SC gap and the lead bandwidth are the largest energy scales, the zero-temperature conductance reads
\begin{equation}
G_{\tilde{q}} = \frac{\frac{2e^2}{h}\,E^2\left(\eta_1^4+\eta_2^4+2\eta_1^2\eta_2^2\cos2\tilde{\chi}\right)}{\left(\frac{E^2-\Delta_\varepsilon^2(\tilde{q})-\left(\pi\nu_l\eta_1\eta_2\sin\tilde{\chi}\right)^2}{\pi\nu_l}\right)^2+E^2(\eta_1^2+\eta_2^2)^2}\,.
\label{eq:G_setup1PF_grounded}
\end{equation}
Here $\nu_l$ is the lead density of states and $E=eV_b$ represents the bias voltage $V_b$. $\Delta_\varepsilon(\tilde{q})=4\varepsilon\cos(4\pi\tilde{q}/3+\phi)$ is the parafermion energy splitting and $\tilde{\chi}=\chi_2-\chi_1$.
When the lead is coupled to a single parafermion ($\eta_2=0$) and the parafermions are not interacting ($\varepsilon=0$), we observe a zero-bias peak for all values of $\tilde{q}$. For $\varepsilon\neq0$ two conductance peaks appear at $E=\pm \Delta_\varepsilon(\tilde{q})$. Moreover, for $\tilde{\chi}=0$ the peaks are quantized at $G=2e^2/h$, analogously to two-Majorana devices \cite{flensberg2010,law2009}. 

\begin{figure}[t]
\centering
\begin{minipage}{0.88\columnwidth}
\hspace{0.17cm}\raisebox{0.1ex}{\includegraphics[width=0.455\columnwidth]{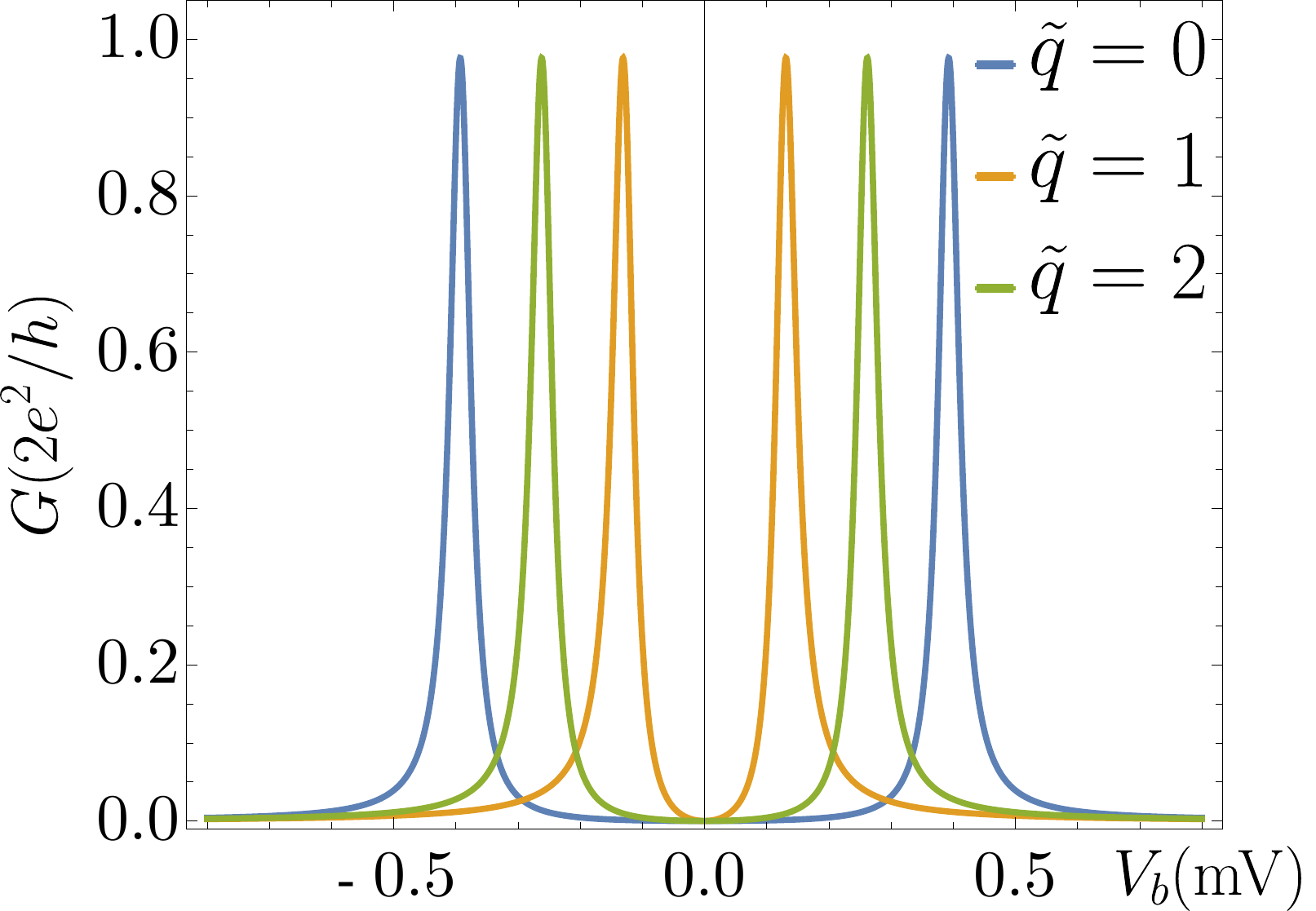}}
\llap{\parbox[b]{7.15cm}{\color{black}(a)\\\rule{0ex}{2.3cm}}}
\hfill
\includegraphics[width=0.48\columnwidth]{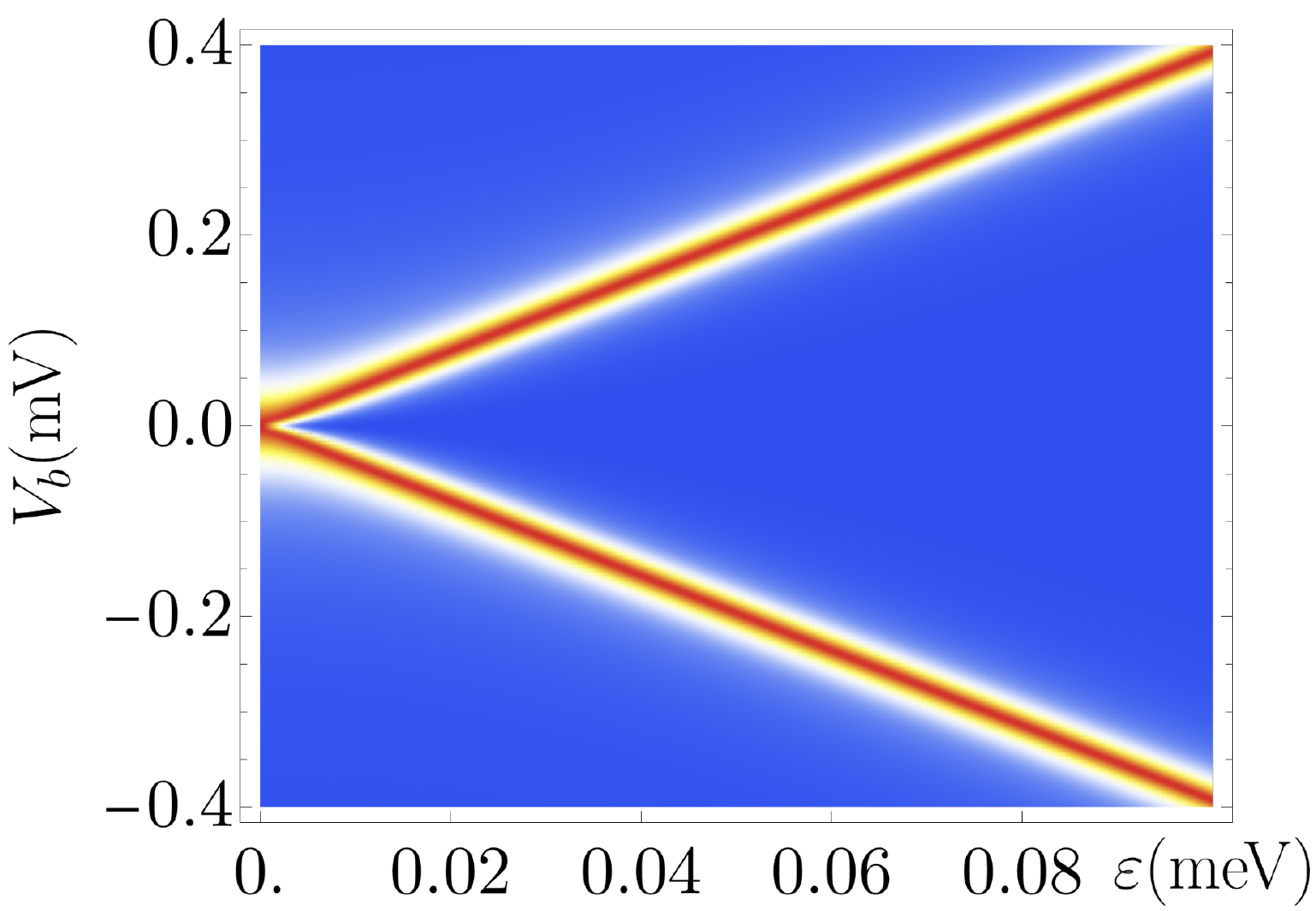}
\llap{\parbox[b]{7.2cm}{\color{black}(b)\\\rule{0ex}{2.3cm}}}
\includegraphics[width=0.48\columnwidth]{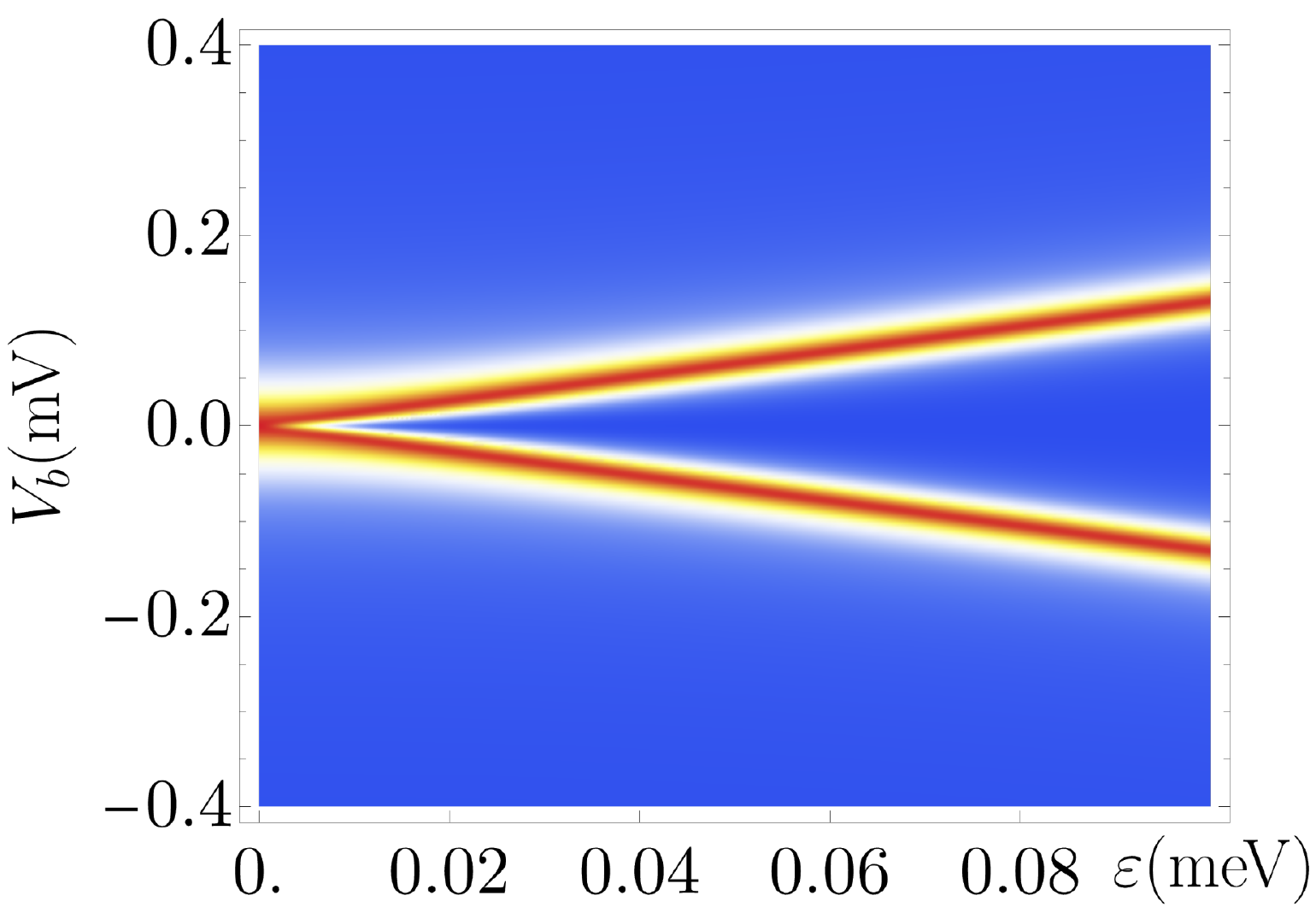}
\llap{\parbox[b]{7.2cm}{\color{black}(c)\\\rule{0ex}{2.3cm}}}
\hfill
\includegraphics[width=0.48\columnwidth]{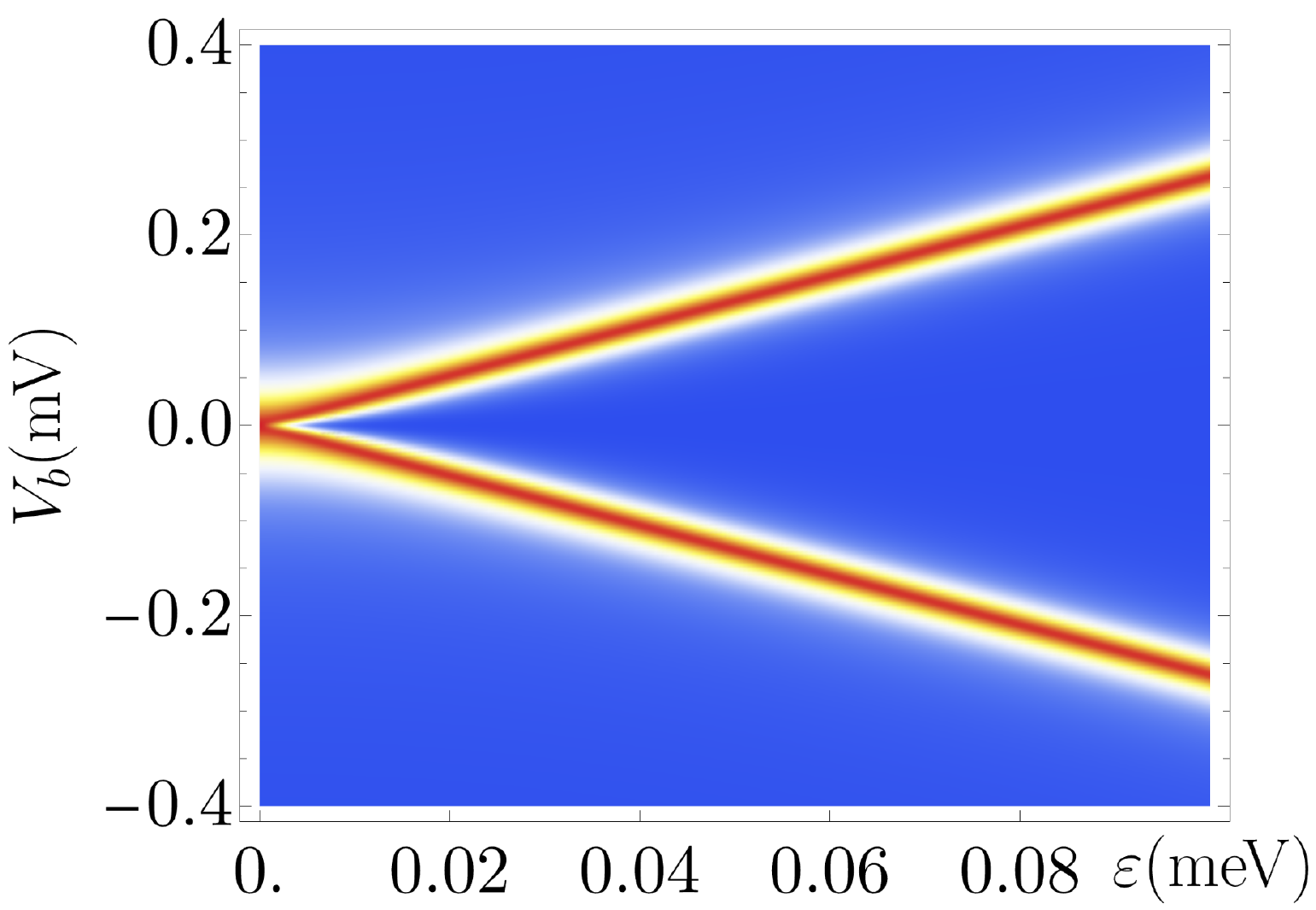}
\llap{\parbox[b]{7.2cm}{\color{black}(d)\\\rule{0ex}{2.3cm}}}
\end{minipage}
\begin{minipage}{0.105\columnwidth}
\includegraphics[width=\textwidth]{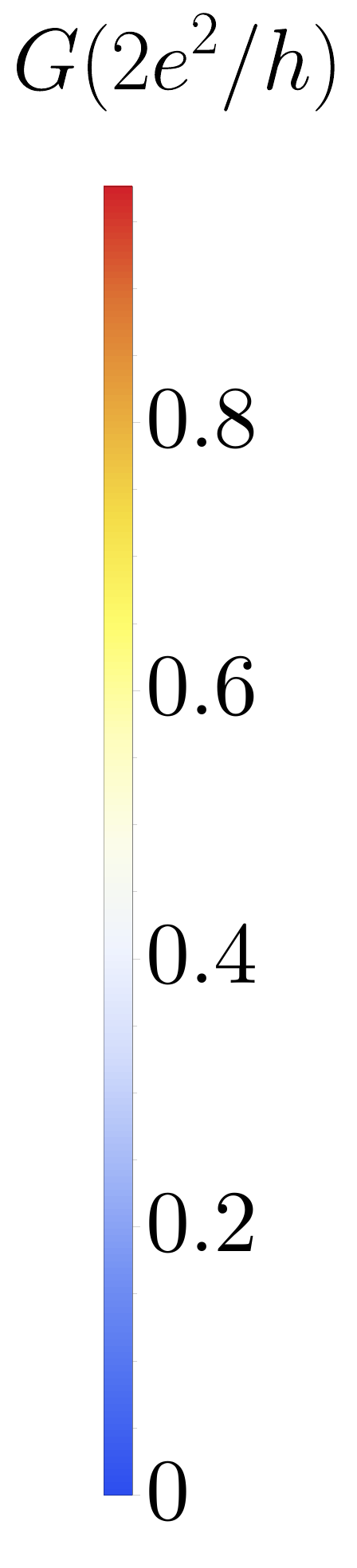}
\end{minipage}
\caption{Conductance of the grounded device as a function of $V_b$ (a) and parafermion overlap $\varepsilon$ (b),(c),(d) for $\nu_l=1.727$ $({\rm  meV})^{-1}$, $\eta_1=0.084$ meV, $\eta_2=0.0082$ meV, and $\tilde{\chi}=0$. (a) Conductance peaks for $\varepsilon=0.1$ meV. (b),(c),(d) System conductance vs $\varepsilon$ for $\tilde{q}=0,1,2$, respectively.}
\label{fig:2terminal}
\end{figure}

This result can be generalized to finite temperature $T$ \cite{suppl}: In Fig.~\ref{fig:2terminal}, choosing typical material parameters and $T= 20$ mK, we show results for the different fractional charges $e\tilde{q}/3$. We choose $\phi=0.19$ for optimal splitting of the degeneracies in $H_{\rm pf}$, and we emphasize that to read the degree of freedom $\tilde{q}$ from a conductance measurement, the parafermion coupling must be sufficiently large, $\varepsilon> T,\eta_1,\eta_2$, such that the different peaks are clearly resolved, as in Fig.~\ref{fig:2terminal}(a).

If the two parafermions are not sufficiently isolated from other fractional modes, decoherence will affect $\tilde{q}$. Consider, for example, events in which the system is poisoned by fractional quasiparticles from the bulk or the external edges of the FQH liquid. If the poisoning rate is sizeable compared with the current measurement time, the stationary state of the parafermion device becomes a statistical mixture of the three values of the parameter $\tilde{q}$, such that the transport readout displays all three peaks in Fig.~\ref{fig:2terminal}(a) at the same time with suitable weights.

In this case, the resulting three-peak differential conductance would be hardly distinguishable from an analogous system hosting three nontopological subgap states, such as Andreev states which may form between the superconductor and the metallic lead. The observation of the conductance patterns in Fig.~\ref{fig:2terminal} is indeed not sufficient to establish the presence of parafermions. For weak poisoning, however, there is a crucial distinction between trivial bound states and parafermion systems. The former can be described by a simple scattering matrix approach and their transport is dictated by a single Landauer-B\"uttiker out-of-equilibrium steady state with a well-defined conductance. There is no possibility of changing their fractional charge for energies below the SC gap. An isolated two-parafermion system, instead, displays three different steady states labelled by $\tilde{q}$. If the poisoning rate is sufficiently weak compared to the current measurement time, the conductance at suitably chosen values of $V_b$ will be affected by a three-state telegraph noise, similar to FQH interferometers \cite{kang2011,rosenow2012}. The corresponding sudden jumps cannot be obtained without fractional subgap states and constitute a strong signature of the quasidegenerate parafermion states, without counterpart in Majorana devices.

The weak poisoning requirement may be hard to meet in experiments. Therefore, we next address complementary devices which, instead, display their main parafermion signatures at stronger poisoning rates and in the truly topological regime $\varepsilon \to 0$.

\begin{figure*}[t]
\raisebox{0.43ex}{\includegraphics[width=0.42\columnwidth]{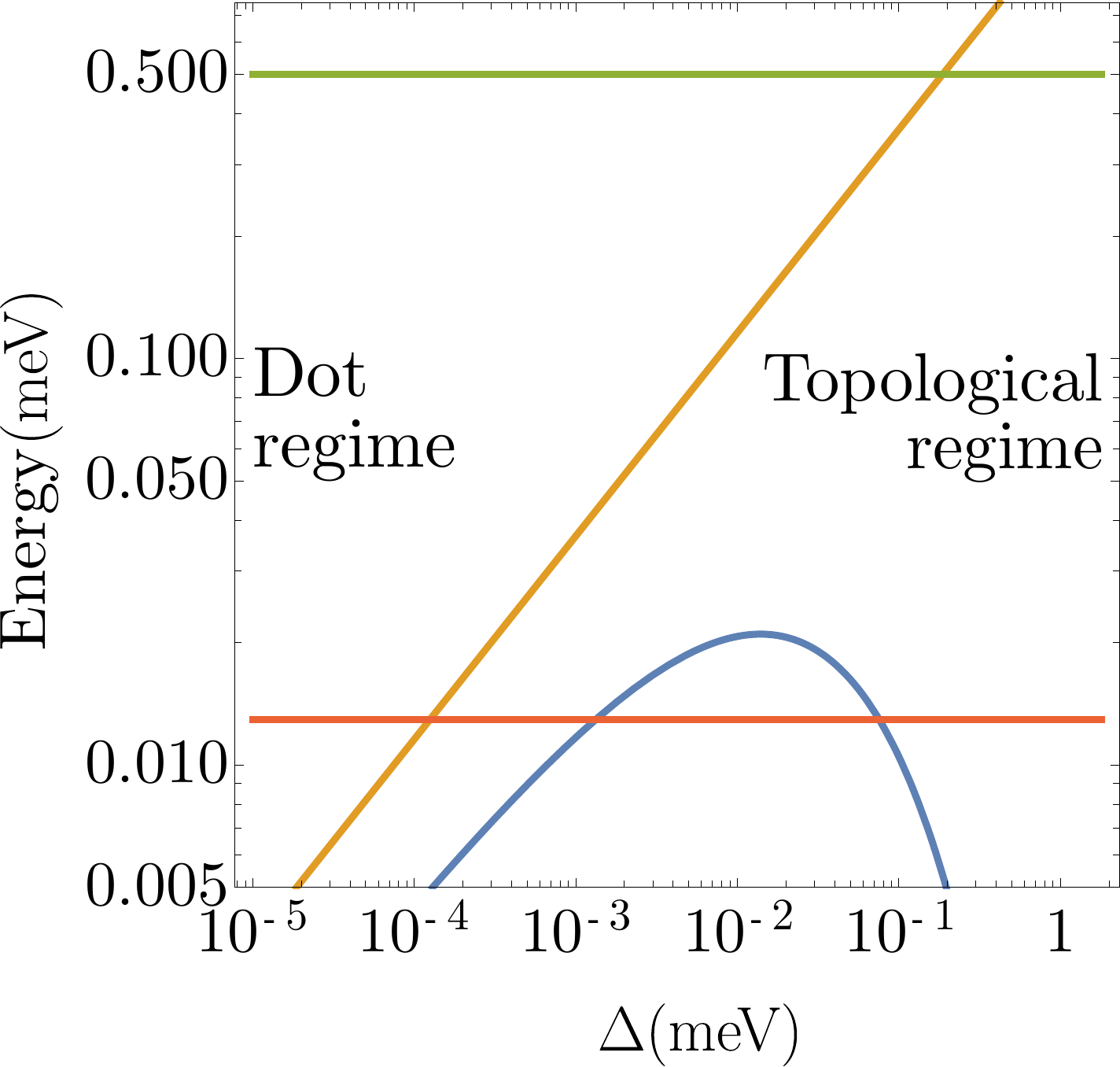}}
\hspace{0.1ex}
\llap{\parbox[b]{6.6cm}{\color{black}(a)\\\rule{0ex}{3.4cm}}}
\includegraphics[width=0.439\columnwidth]{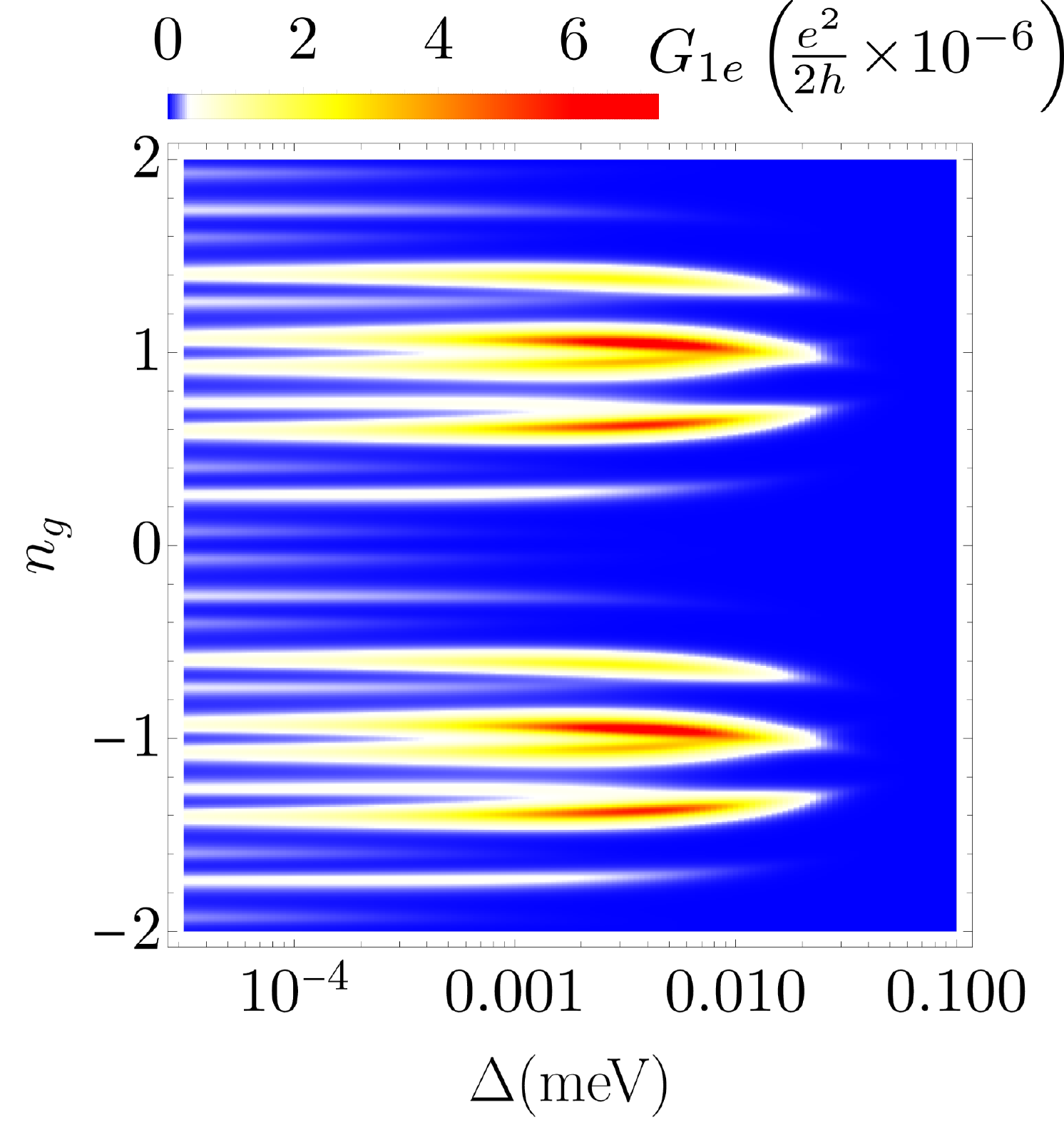}
\hspace{-1.7ex}
\llap{\parbox[b]{7.cm}{\color{black}(b)\\\rule{0ex}{3.4cm}}}
\raisebox{0.2ex}{\includegraphics[width=0.412\columnwidth]{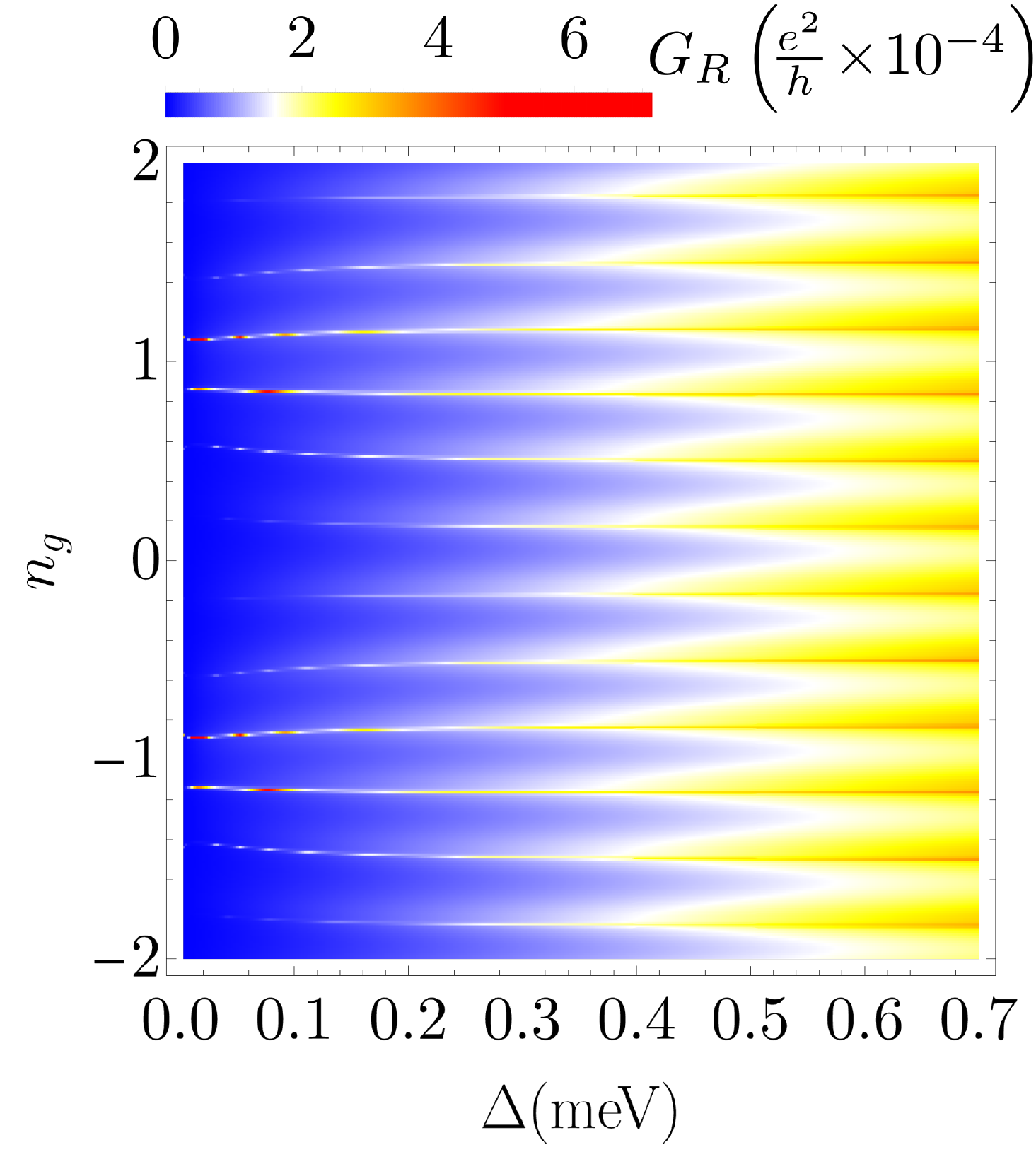}}
\llap{\parbox[b]{6.8cm}{\color{black}(c)\\\rule{0ex}{3.4cm}}}
\raisebox{1.1ex}{\includegraphics[width=0.345\linewidth,trim={4.7cm 10.7cm 4.8cm 10.9cm},clip]{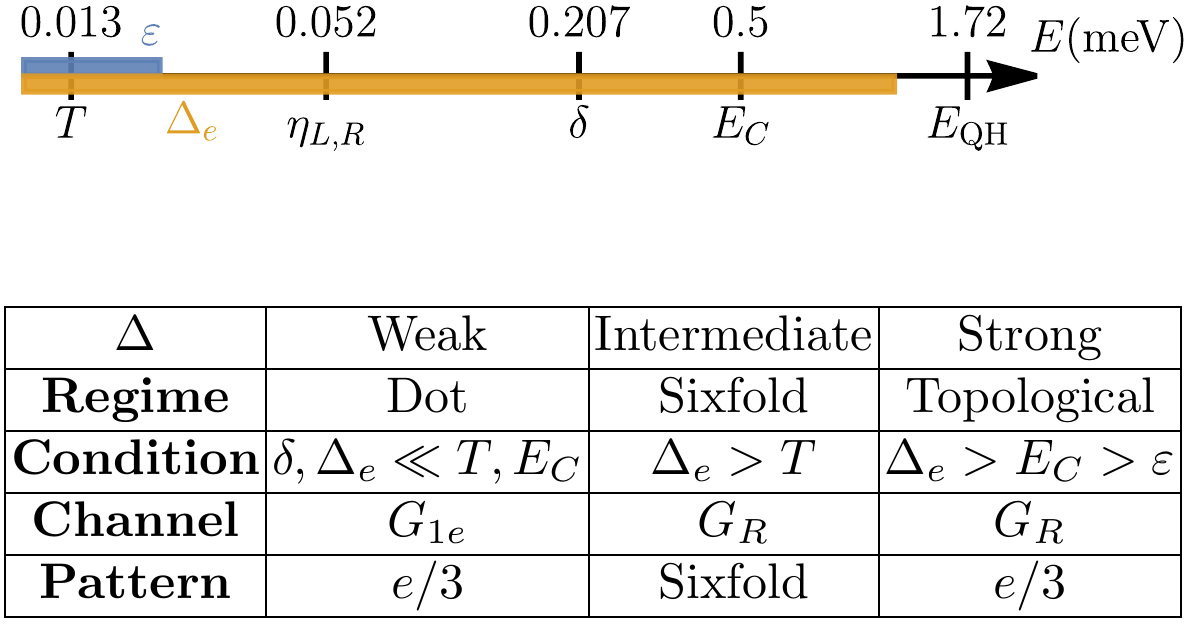}}
\llap{\parbox[b]{12.35cm}{\color{black}(e)\\\rule{0ex}{1.85cm}}}
\llap{\parbox[b]{12.5cm}{\color{black}(d)\\\rule{0ex}{3.4cm}}}
\caption{Blockaded device coupled to two metallic leads with strengths $\eta_L=\eta_R=0.052$ meV  \cite{suppl}. (a) Based on realistic parameters ($E_{\rm QH}=1.72$ meV, $v=10^5$ m/s, $L=1$ $\mu$m), $\varepsilon$ (blue) and $\Delta_e$ (orange) are compared with the charging energy $E_C = 0.5$ meV (green) and temperature $T=0.15$ K (red). (b) Sequential tunnelling zero-bias conductance $G_{1e}$ as a function of $\Delta$ and $n_g$ [from Eq.~\eqref{eq:Gtot} and \cite{suppl}]. (c) Resonant parafermion-mediated zero-bias conductance $G_R$ [Eq.~\eqref{eq:Gtot} and \cite{suppl}]. This is typically the dominant contribution from moderate to large $\Delta$. A sixfold pattern is observed for $\Delta\lesssim 0.3$ meV and evolves into an $e/3$-periodic pattern with increasing $\Delta$. (d) Log scale summary of energies adopted in (b) and (c), including the dot energy level spacing $\delta$. Ranges of $\varepsilon$ and $\Delta_e$ are indicated by blue and orange. (e) Main properties of the discussed transport regimes.} \label{fig:float}
\end{figure*}

\textit{The Coulomb blockaded device.}---For intermediate fractional quasiparticle poisoning, different signatures of the parafermions can be sought in the transport features of Coulomb blockaded SC islands. In the past years, several tunnelling spectroscopy experiments have examined the onset of MZM through the two-terminal conductance of analogous devices \cite{albrecht2016,albrecht2017,kanne2021,sole2020}. Their topological phases can be distinguished by a different periodicity of the zero-bias conductance peaks as a function of the induced charge $n_g$ \cite{fu2010,egger12,vanheck2016}: $2e$ and $1e$ periodic patterns are observed in the trivial and topological phases, respectively. 

In the following, we analyse the two-terminal conductance through a setup in which the NbN finger is floating and constitutes a SC island with strong charging energy. Its induced charge $n_g$ can be varied by a top gate voltage $V_g$ [Fig.~\ref{fig:2pfsetup}(c)] and we consider two noninteracting metallic leads in proximity to the parafermions. The Hamiltonian of the floating SC island is modelled by
\begin{multline}
H_{\rm SC}(N_{\rm C},N_e,N',q,n_g) = H_{\rm pf}(q) +  N_e \Delta_e + N' \Delta_{e/3} \\ + E_C \left(2N_{\rm C} + N_e + N'/3+q/3-n_g\right)^2.
\label{eq:H_SC}
\end{multline}
Here we label the system states by the occupation numbers $\ket{N_{\rm C},N_e,N',q}$, where $N_{\rm C}$ refers to Cooper pairs in the SC island, $N_e$ to quasielectron excitations in the device, $N'$ to fractional quasiparticle excitations, and $eq/3$ is the charge of the localized parafermions as before.
$H_{\rm SC}$ is determined by the following energy scales: the island charging energy $E_C$, the parafermion coupling $\varepsilon$, the energy gap of a quasielectron in the paired FQH edge modes $\Delta_e$, and the energy gap for an $e/3$ quasiparticle excitation $\Delta_{e/3}$. To estimate $\varepsilon,\,\Delta_e$, and $\Delta_{e/3}$, we model the system as a Luttinger liquid, describing the counterpropagating fractional edge modes beneath the SC finger \cite{shtengel2012}. Their proximity-induced SC pairing is treated at a mean-field level by introducing a crossed Andreev interaction $\Delta$ between the chiral modes \cite{suppl}, which is proportional to the SC gap of the NbN finger; it decays with its width \cite{kim2017} and decreases with magnetic field. The fractional quasiparticle gap is estimated through a semiclassical analysis: these quasiparticles can be described as solitons in a sine-Gordon model \cite{burnell2016,mussardo,suppl} and their mass results in $\Delta_{e/3} = \sqrt{8 \Delta E_{\rm QH} / 3 \pi^3}$ as a function of $\Delta$ and the bulk gap $E_{\rm QH}$ of the FQH state with typical value $E_{\rm QH} \sim 2$ meV \cite{suppl,bolotin2009}. The results in Ref.~\cite{kim2020} suggest a rough estimate $\Delta \lesssim 1$ meV. In the following, we  adopt $\Delta$ as the main parameter to distinguish the system regimes and we exploit a simplified low-energy description. We approximate the behaviour of both fractional and electron quasiparticles with noninteracting dynamics \cite{suppl} and we minimize the quasielectron gap by $\Delta_e = 3\Delta_{e/3}$. Finally, the parafermion energy splitting $\varepsilon$ is \cite{burnell2016}
\begin{equation}
\varepsilon = \sqrt{{\pi}/{2}}\,\Delta_{e/3} \exp\left[-\sqrt{{2\pi}/{3}} \, \left(L\Delta_{e/3}/{\hbar v}\right)\right]\,,
\label{eq:PFsplit}
\end{equation}
where $v$ the velocity of the chiral edge modes [see Fig.~\ref{fig:float}(a)].

Analogously to Majorana devices \cite{vanheck2016,albrecht2017,hansen2018}, transport across the system can be modelled by considering two main processes: (i) resonant tunnelling of electrons mediated by parafermions, corresponding to electron teleportation mediated by the $\gamma_i=\alpha_i^3$ MZM \cite{fu2010}, and (ii) single-electron incoherent sequential tunnelling across gapped chiral states. A third process---sequential Andreev tunnelling of Cooper pairs---is of less practical relevance and is discussed in the SM \cite{suppl}.

Similarly to long grounded systems, the resonant tunnelling (i) accounts for transitions between $q$ and $q+3 \mod 3$ caused by the couplings of neighbouring lead and parafermion [analogous to Eq.~\eqref{coupling} setting  $\eta_2=0$ \cite{suppl}]. Concerning the sequential tunnelling (ii), we assume instead that electrons from the leads hop in and out of the island causing transitions between $N_e=0$ and $N_e=1$. We neglect scattering processes of the quasielectrons into fractional particles \cite{suppl}, such that $N'$ is conserved. We phenomenologically capture the quasiparticle poisoning from the external environment by considering a thermal equilibrium distribution of the states $\ket{N_{\rm C},N_e,N',q}$ at temperature $T \ll E_C$. Hence, we estimate the total differential conductance by
\begin{multline}
G= G_R + G_{1e}= \hspace{-1em} \sum_{N_{\rm C},N_e,N',q} \hspace{-1em} Z^{-1}\e^{-H_{\rm SC}(N_{\rm C},N_e,N',q,n_g)/T} \\
 \times \sum_{a=R,1e} \tilde{G}_{a}\left(n_g - \left[2N_{\rm C}+N_e+(N'+q)/3\right]\right),
 \label{eq:Gtot}
\end{multline}
where the indices $R$ and $1e$ represent resonant and sequential tunnelling, respectively, and $Z$ is the partition function. 
The conductances $\tilde{G}_a$ are estimated based on rate equations, analogously to the Majorana devices \cite{vanheck2016} (see the SM \cite{suppl} for details), and the resulting $G_a$ are exemplified in Figs.~\ref{fig:float}(b) and \ref{fig:float}(c).

For a realistic parameter choice, we can distinguish three main conductance patterns as a function of the SC pairing $\Delta$, which we label as quantum dot, sixfold, and topological regimes [see table in Fig.~\ref{fig:float}(e)]. For $\Delta \to 0$, the gaps $\Delta_{e/3},\,\Delta_e$, and $\varepsilon$ vanish, thus the system behaves as a blockaded dot for fractional quasiparticles \cite{goldman1995,mills2020}. The electron sequential tunnelling dominates and results in a complex set of zero-bias conductance peaks alternating with a $2e$ periodicity in the induced charge $n_g$. The location of these peaks
depends on a further energy scale, $\delta = vh/2L$. In the example depicted in Fig.~\ref{fig:float}(b), the dot energy level spacing $\delta$ is comparable with $E_C$, thus yielding twelve irregular peaks in each $2e$ period for $\Delta \to 0$. Systems with $\delta, \Delta_e \ll E_C$, instead, would display an $e/3$ periodic pattern (not shown).

For intermediate pairing, the resonant tunnelling becomes relevant and the most common zero-bias pattern displays only six dominant peaks repeating with $2e$ periodicity: We call this regime \textit{sixfold} [left side of Fig.~\ref{fig:float}(c), for $0.005$ meV $\lesssim \Delta \lesssim 0.3$ meV]. Such pattern emerges when $\Delta_e$ becomes larger than $T$.

Finally, for strong induced pairing $\left(\Delta \to E_{\rm QH}\right)$ in sufficiently long islands, the parafermion splitting drops, $\varepsilon \ll E_C < \Delta_e$ [Eq.~\eqref{eq:PFsplit} and Fig.~\ref{fig:float}(a)]. Therefore, the system is deeply in the topological regime and the zero-energy parafermions yield zero-bias $G_R$ peaks repeating with a characteristic $e/3$ periodicity [Fig.~\ref{fig:float}(c), right side], analogously to the Majorana-mediated electron teleportation \cite{fu2010,vanheck2016,rosenow2021}.

In conclusion, for typical experimental parameters ($L=1$ $\mu$m, $E_C=0.5$ meV), this $e/3$ periodicity signals the onset of the topological phase with strongly localized parafermions. Only for devices with negligible quasielectron excitation energy (dot regime with $\delta,\Delta_e \ll T < E_C$), an additional $e/3$-periodic pattern may appear [table in Fig.~\ref{fig:float}(e)].

\textit{Conclusions.}---We showed that parafermions in FQH setups with induced superconductivity \cite{kim2020} can be investigated through electronic transport measurements. In suitable two-terminal devices they give rise to conductance peaks analogously to MZM. In contrast with MZM, however, the ground states of isolated pairs
of interacting parafermions can be distinguished through a current readout. The different values of their shared fractional charge yield different low-energy resonances in the conductance between a metallic electrode (tunnel coupled to the parafermions) and the SC background. This distinguishes our setup from devices characterized by the transport of fractional quasiparticles \cite{barkeshli2014b,kim2017b,egger2018,buccheri2018,mazza2018,oreg2020,klinovaja2020}.
In the presence of weak quasiparticle poisoning we expect to observe a three-state telegraph noise, discriminating between parafermion and trivial electronic subgap states.

Complementary signatures of the parafermions are obtained from the analysis of blockaded devices. We studied the two-terminal conductance across a floating fractional superconductor as a function of its induced charge. For intermediate SC pairing and low temperature, the zero-bias conductance is characterized by a six-peak pattern repeating with $2e$ periodicity. The zero-bias peaks evolve towards an $e/3$-periodic pattern for strongly localized parafermions (similarly to fractional quasiparticle transport \cite{kim2017b}). 

The electronic tunnelling spectroscopy we presented can be generalized to devices including additional leads or quantum dots, as recently proposed \cite{teixeira2021} for non-topological $\mathbb{Z}_4$ parafermions \cite{mazza2018b,calzona2018,chew2018}, and it can be integrated with additional fractional quasiparticle elements to develop novel topological quantum computation platforms based on parafermions.

\textit{Acknowledgements.}---We warmly thank Y. Gefen, F. Buccheri, A. C. C. Drachmann, M. Leijnse, M. Wauters, and A. B. Hellenes for useful discussions and K. Snizhko for helpful comments on a preliminary version of this manuscript and insightful observations. I.~E.~N.~and M.~B.~are supported by the Villum foundation (research Grant No.~25310). We acknowledge support from the Deutsche Forschungsgemeinschaft (DFG) project Grant No. 277101999 within the CRC network TR 183 (subproject C01), as well as Germany's Excellence Strategy Cluster of Excellence Matter and Light for Quantum Computing (ML4Q) EXC 2004/1 390534769 and Normalverfahren Projektnummer EG 96-13/1. This project also received funding from the European Research Council (ERC) under the European Union’s Horizon 2020 research and innovation program under Grant Agreement No.~856526, from the Danish National Research Foundation, and the Danish Council for Independent Research $|$ Natural Sciences.

\section{Supplemental material: The scattering matrix of the grounded N-SC setup} 
\begin{figure}
\centering
\includegraphics[width=0.9\linewidth,clip,trim={2cm 0 3cm 0}]{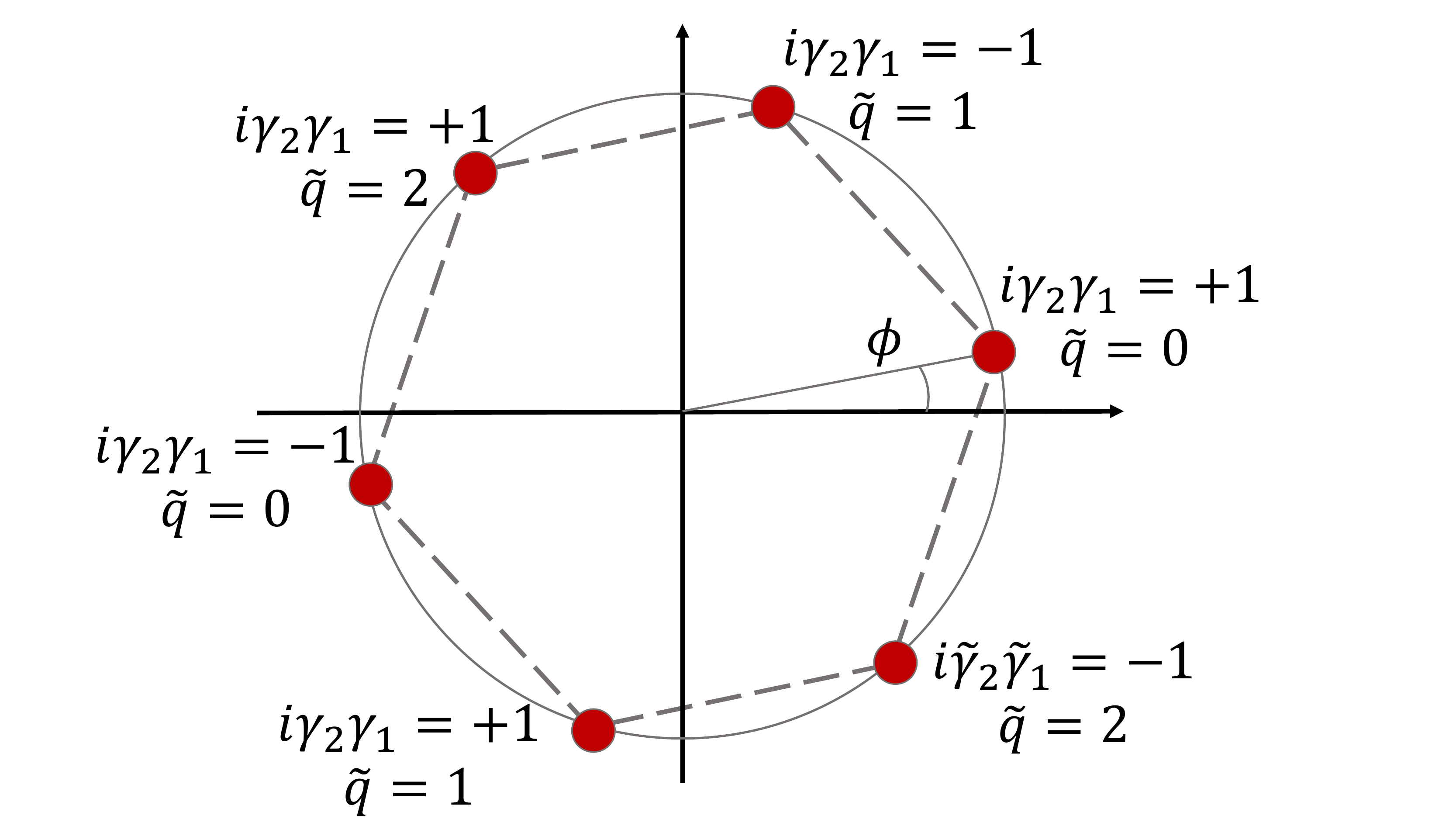}
\caption{Illustration of the six energy levels of the overlapping parafermions $\alpha_1$ and $\alpha_2$, and of how a finite $\phi$ can shift these. The projection onto the horizontal axis is proportional to this degeneracy splitting. For the optimal value $\phi=\arctan(1/\sqrt{27})$ the three energies $|-2\varepsilon\cos(4\pi\tilde{q}/3+\phi)|$ for $\tilde{q}=0,1,2$ are maximally separated. For $\phi=0$ the sectors $\tilde{q}=1,2$ are degenerate.}
\label{fig:Z6energy}
\end{figure}

With the Weidenm\"uller formula \cite{aleiner2002} we can calculate the conductance between a metallic and a SC lead mediated by two parafermions described by $H_{\rm pf}$ [Eq.~\eqref{eq:H_pf}]. The use of this formula requires the lead bandwidth to be the largest energy scale in the system, and that the lead density of states $\nu_l$ can be assumed constant in the energy range defined by the bias voltage. We introduce the matrix $W$ \cite{anna2020} such that the coupling Hamiltonian in Eq.~\eqref{coupling} reads
\begin{equation}
H_c = \underline{c}^\dagger W \underline{l} + \rm{H.c.}\,,
\end{equation}
where $\underline{l}^\dagger=(l^\dagger,l)$, $\underline{c}^\dagger=(c^\dagger,c)$, and $c=(\gamma_1+i\gamma_2)/2$ is the annihilation operator of the effective subgap state describing the fermionic part of $H_{\rm pf}$, with $\gamma_i=\alpha_i^3$. Explicitly, the $W$ matrix thus reads
\begin{equation}
W=\frac{1}{2}\begin{pmatrix}
i\eta_1-\eta_2 \e^{i\tilde{\chi}} & i\eta_1-\eta_2 \e^{-i\tilde{\chi}} \\
i\eta_1+\eta_2 \e^{i\tilde{\chi}} & i\eta_1+\eta_2 \e^{-i\tilde{\chi}}
\end{pmatrix}\,.
\end{equation}
The corresponding scattering matrix results \cite{aleiner2002}
\begin{equation} \label{weiden}
S(E)= \Id -i 2\pi \nu_l W^{\dag} \frac{1}{E-H_{\rm pf}^{\rm BdG}(\tilde{q})+i\pi\nu_l W W^{\dag}}W\,.
\end{equation}
Here $H_{\rm pf}^{\rm BdG}(\tilde{q})=4\varepsilon\cos(4\pi\tilde{q}/3+\phi)\tau_z$ is the Bogoliubov-de Gennes Hamiltonian describing the energy splitting of the fermionic part of $H_{\rm pf}$ which depends on the fractional charge $e\tilde{q}/3$, and $\tau_z$ is the third Pauli matrix in electron-hole space. The fractional charge number $\tilde{q}=0,1,2$ corresponds in turn to the following expression in terms of the parafermion operators, derived by the convention adopted in the main text, $\e^{i\pi q/3} = i\gamma_2\gamma_1\e^{i\pi 4\tilde{q}/3}$:
\begin{equation}
\e^{i4\pi\tilde{q}/3}=\e^{-i\pi/3}(\alpha_2^\dagger\alpha_1)^2= \e^{-i2\pi q/3} \,.
\end{equation}
The scattering matrix $S$ is two-dimensional and by the Landauer-B\"uttiker formalism and unitarity of $S$, the zero-temperature expression for the differential conductance is $G_{\tilde{q}}(E)=\frac{2e^2}{h}|R_{he}|^2$, where $R_{he}$ is the off-diagonal component of $S$. The resulting conductance $G_{\tilde{q}}(E)$ is given in Eq.~\eqref{eq:G_setup1PF_grounded}. The peak broadness is set by the coupling strengths $\eta_{1,2}$ and similar to that of non-interacting subgap states such as Andreev bound states. Notice that the scattering matrix is independent of the length $L_c$ of the contact since $\eta_i\propto1/\sqrt{L_c}$ and $\nu_l\propto L_c$. The value of $\nu_l$ in Fig.~\ref{fig:2terminal} is set by assuming gold leads with Fermi velocity $v_{\rm F}=1.4\times10^6$ m/s \cite{ashcroft} and $L_c=5$ $\mu$m. Considering a lead at temperature $T$ and voltage $V_b$, the finite-temperature conductance is $G_{\tilde{q}}(V_b,T)=\int_{-\omega_D}^{\omega_D}\! d\omega \, G_{\tilde{q}}(\omega)(-n_{\text{F}}'(\omega))$, where $n_{\text{F}}(\omega)=(\e^{(\omega-eV_b)/(k_{\rm B} T)}+1)^{-1}$ is the Fermi distribution function in the lead, $k_{\rm B}$ the Boltzmann constant, and $\omega_D$ an appropriate cut-off frequency. This is plotted in Fig.~\ref{fig:2terminal} for the different $\tilde{q}$ where, as for the rest of the paper, the phase $\phi$ is set to $\arctan(1/\sqrt{27})\approx0.19$ to optimize the distance between the peaks. Fig.~\ref{fig:Z6energy} illustrates the effect of $\phi$ on the six low-energy states of $H_{\rm pf} = -i2\varepsilon \gamma_2\gamma_1 \cos(4\pi\tilde{q}/3+\phi)$.

Finally, we emphasize that the detection of these conductance peaks, as well as other spectroscopy techniques, is necessary to establish the presence of subgap states, but not sufficient to confirm their topological origin. In the case of Majorana modes, complementary approaches concerning their braiding, non-local, or thermodynamic properties have been theoretically proposed \cite{sela2019,kasahara2018,crepel2019,beenakker2020,flensberg2021}.

\section{Supplemental material: General form for the lead-parafermion coupling} \label{app:coupling}
\begin{figure}[t]
\centering
\begin{minipage}{0.88\columnwidth}
	\hspace*{1.3ex}\raisebox{0.2ex}{\includegraphics[width=0.45\columnwidth]{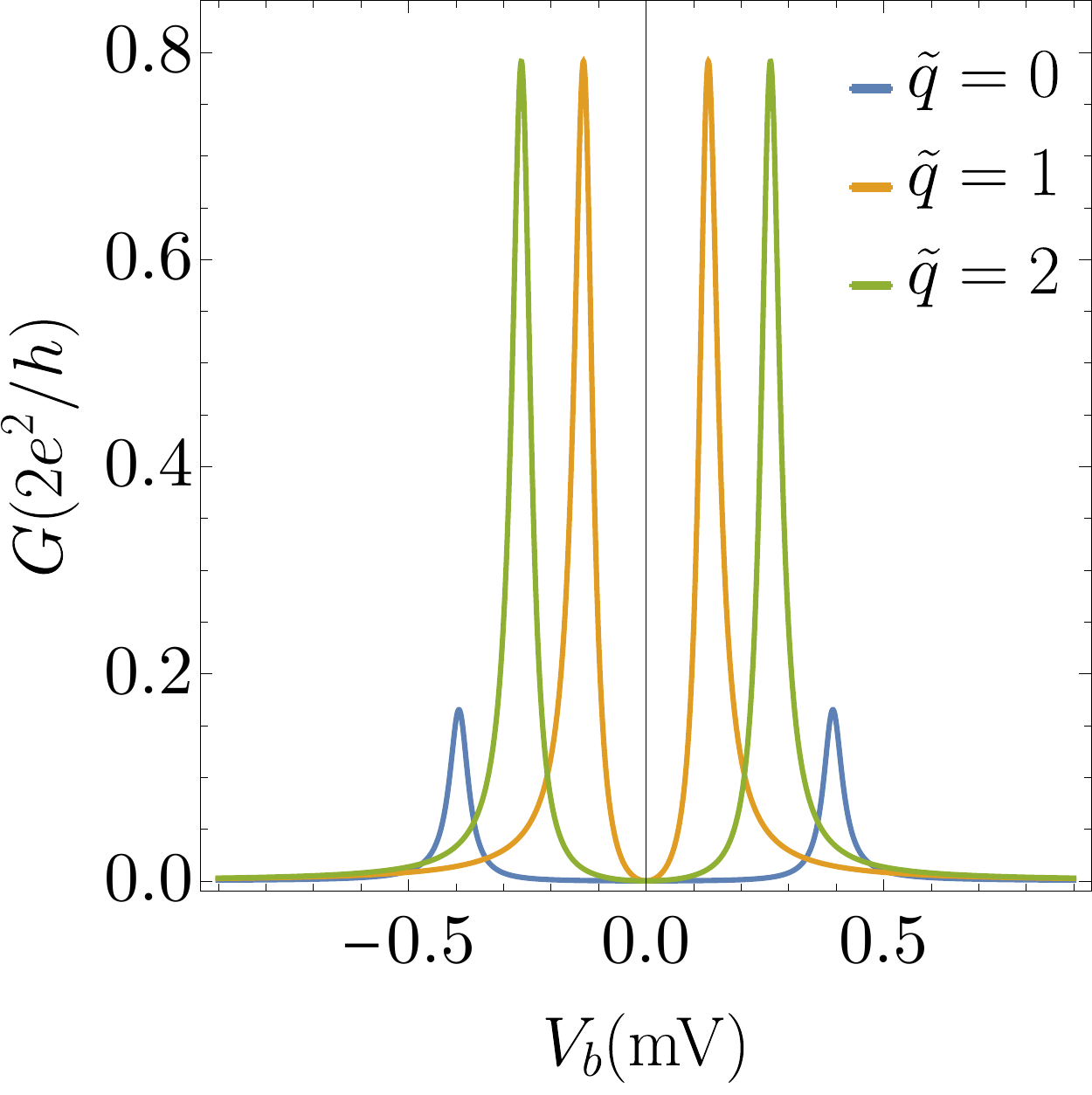}}
	\llap{\parbox[b]{7.0cm}{\color{black}(a)\\\rule{0ex}{3.35cm}}}
	\hfill
	\includegraphics[width=0.48\columnwidth]{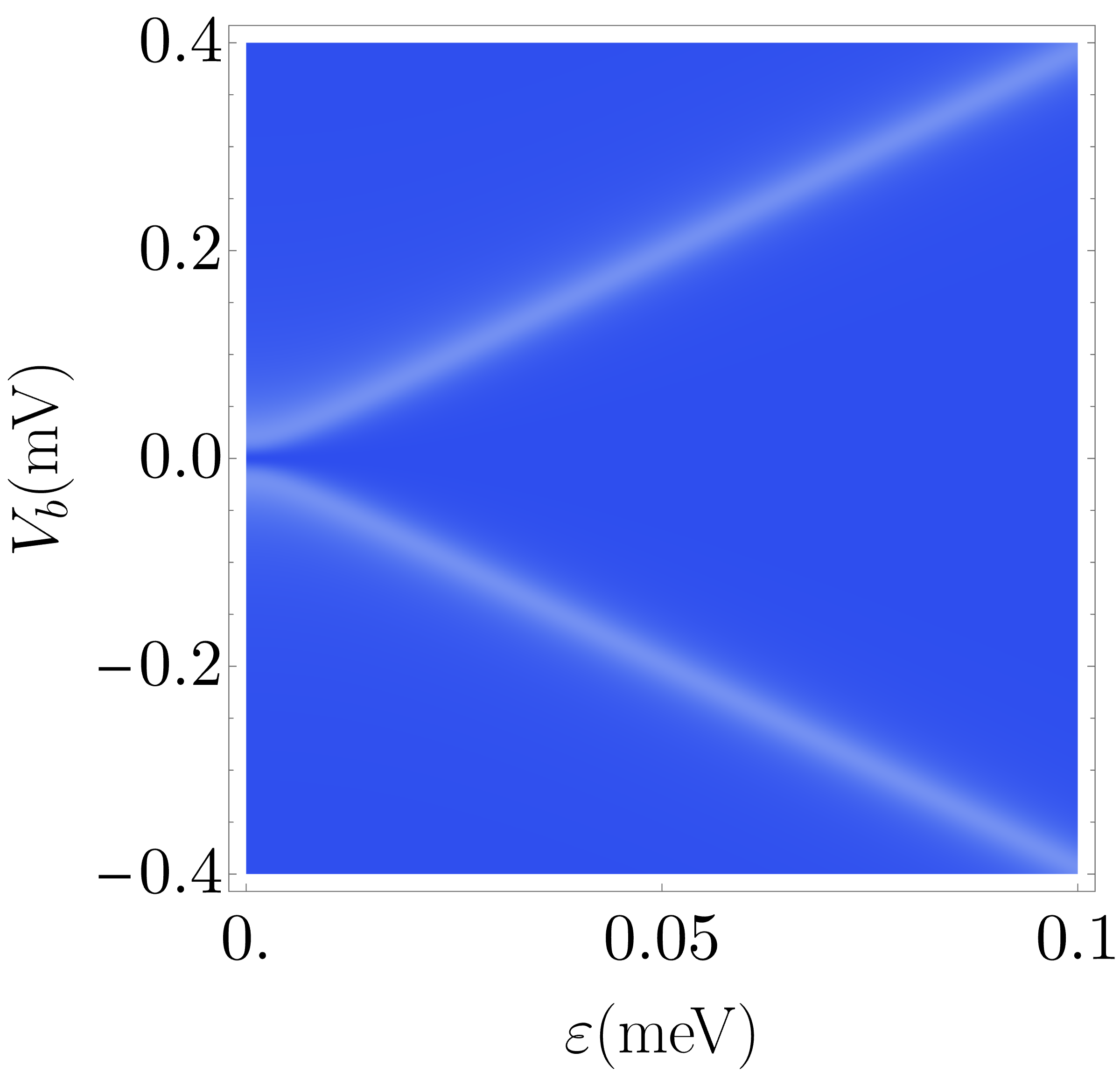}
	\llap{\parbox[b]{7.1cm}{\color{black}(b)\\\rule{0ex}{3.35cm}}}
	\includegraphics[width=0.48\columnwidth]{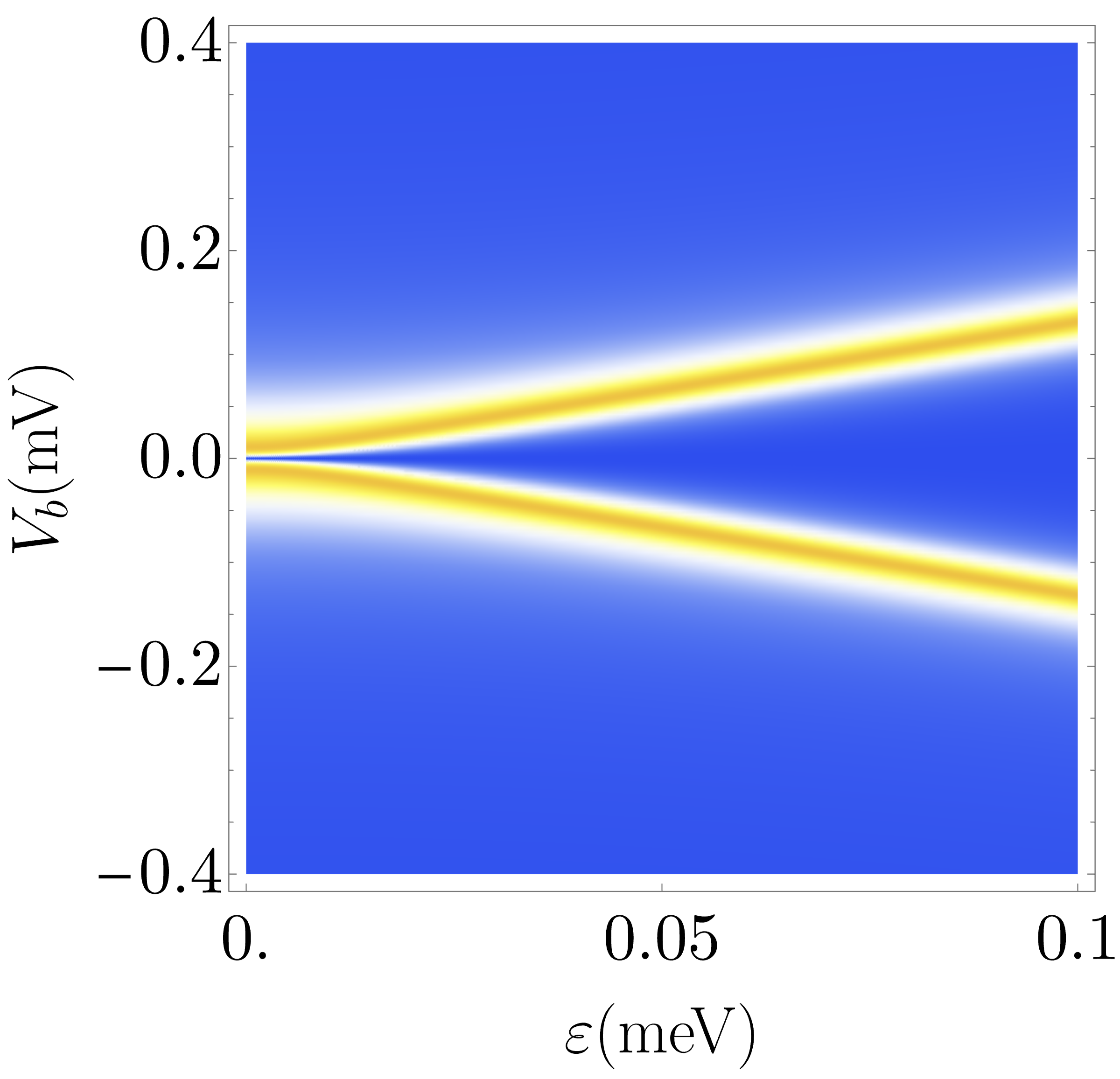}
	\llap{\parbox[b]{7.1cm}{\color{black}(c)\\\rule{0ex}{3.35cm}}}
	\hfill
	\includegraphics[width=0.48\columnwidth]{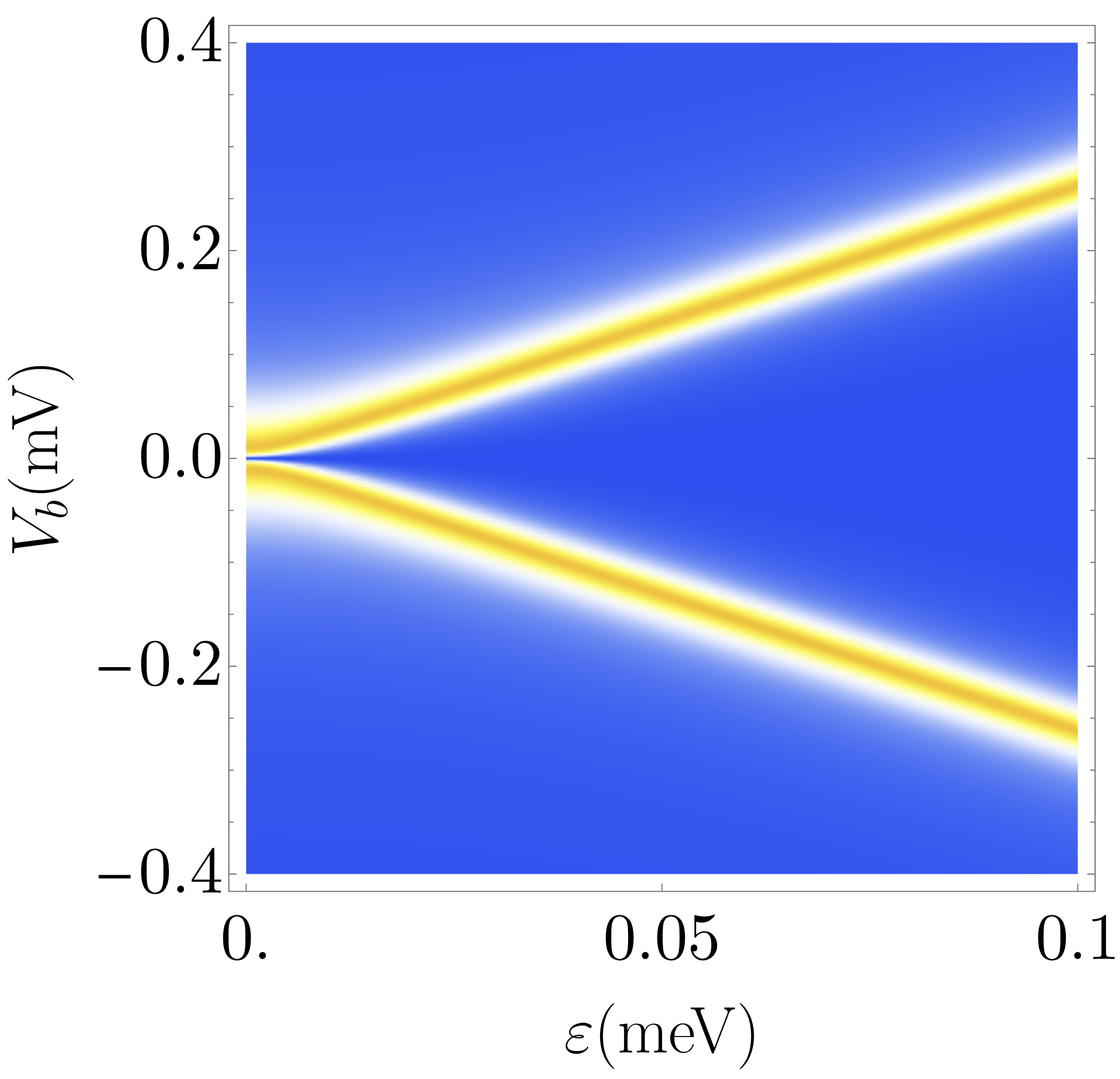}
	\llap{\parbox[b]{7.1cm}{\color{black}(d)\\\rule{0ex}{3.35cm}}}
\end{minipage}
\begin{minipage}{0.1\columnwidth}
	\includegraphics[width=\textwidth]{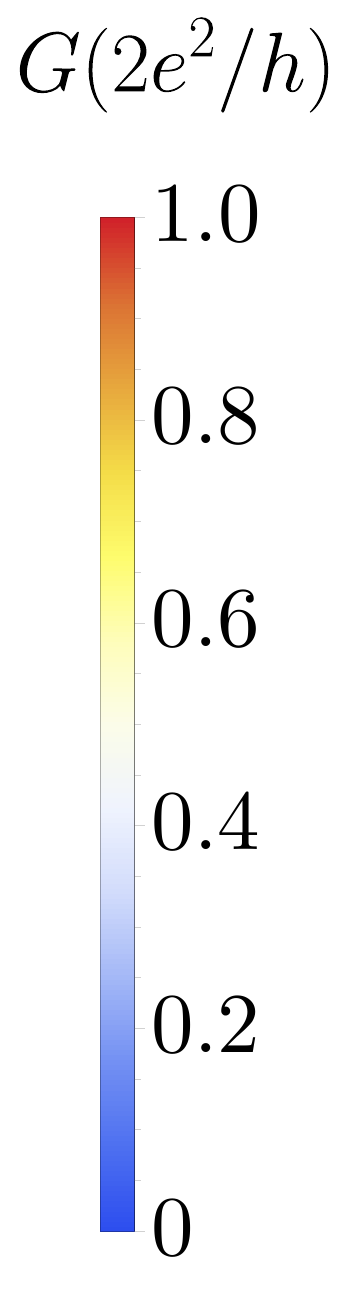}
\end{minipage}
	\caption{Conductance of the grounded device in Figs.~\ref{fig:2pfsetup}(a) and \ref{fig:2pfsetup}(b) as a function of voltage bias $V_b$ and parafermion overlap $\varepsilon$ for the more general coupling Hamiltonian in Eq.~\eqref{coupling2}. For comparison with the results in Fig.~\ref{fig:2terminal} we keep $\nu_l=1.727$ (meV)$^{-1}$, $\phi=0.19$, and $\kappa_0=\eta_1=0.084$ meV with the other couplings given by:  $\kappa_1=\kappa_5=\kappa_0/3, \kappa_2=\kappa_4=\kappa_0/9$, and $\kappa_3=\kappa_0/27$. All phases $\chi_n$ are set to 0. (a) Conductance peaks for $\varepsilon=0.1$ meV. (b),(c),(d) Conductance for $\tilde{q}=0,1,2$, respectively, with values as indicated on the colour scale to the right.}
	\label{fig:2terminal_gen}
\end{figure}

Based on the introduction of the Majorana operators $\gamma_j = \alpha_j^3$, let us consider the most general form for the coupling between a lead and two parafermions that assumes a quadratic form in the fermionic degrees of freedom. We require that all coherent processes coupling the lead with the parafermions vary the parafermion charge by $\pm e$. The most general coupling fulfilling these requirements is
\begin{equation}
 \label{coupling2}
H_c = i \sum_{n=0}^5 \kappa_n \e^{i\chi_n} \alpha_1^{3-n} \alpha_2^n l + {\rm H.c.}\,,
\end{equation}
where we consider the amplitude $\kappa_n$ to be real and assume the parafermionic modes resulting from the FQH edges to be fully polarized in the direction of the strong external magnetic field. Hence we consider only one spin species in the leads to be involved in the tunnelling. We observe that
\begin{multline}
i\alpha_1^3 \alpha_1^{-n} \alpha_2^n = i \gamma_1 \e^{-i\frac{\pi}{3}\frac{n(n-1)}{2}} \left(\alpha_1^\dag \alpha_2\right)^n \\
= i \gamma_1 \e^{-i \frac{\pi}{3}\frac{n^2}{2}}\e^{i\pi n q/3}
= i \gamma_1 \left(i\gamma_2\gamma_1\right)^n \e^{-i\frac{\pi n^2}{6}}\e^{i\pi n 4 \tilde{q}/3}\,,
\end{multline}
where we used the convention presented in the main text and the previous section for the definition of the fractional charge  $\tilde{q}/3$. Therefore we get
\begin{multline}
H_c = \sum_{n \, \textrm{even}} i \kappa_n \gamma_1 \left[l \e^{i\left(\chi_n -\frac{\pi n^2}{6} +\frac{4 \pi}{3} n \tilde{q}\right)} + {\rm H.c.}\right] \\
+ \sum_{n \, \textrm{odd}} i \kappa_n \gamma_2 \left[-i l \e^{i\left(\chi_n -\frac{\pi n^2}{6} +\frac{4 \pi}{3} n \tilde{q}\right)} + {\rm H.c.}\right]\,.
\label{eq:gen_coupling}
\end{multline}
The terms $n=0$ and $n=3$ are independent of $\tilde{q}$ and match the simplified case presented in the main text. Instead, the other terms introduce phases that include $\tilde{q}$. Consequently, the conductance will depend on the fractional charge $\tilde{q}/3$ not only through the parafermion level splitting, but also via the coupling Hamiltonian. We can write the Hamiltonian in Eq.~\eqref{eq:gen_coupling} as in Eq.~\eqref{coupling} with effective strengths and phases $\eta_i^{\text{eff}}$ and $\chi_i^{\text{eff}}$ determined by $\tilde{q}$ and the parameters $\{\kappa_n,\chi_n\}$. The conductance is thus expressed as in Eq.~\eqref{eq:G_setup1PF_grounded} by replacing $\eta_i$ and $\tilde{\chi}$ with those effective parameters. The conductance quantization can then no longer be true for all $\tilde{q}$ sectors since these have different $\tilde{\chi}^{\rm eff}(\tilde{q})=\chi_2^{\text{eff}}-\chi_1^{\text{eff}}$ and will therefore have different peak heights, even when the couplings are originally real, $\chi_n=0$ in Eq.~\eqref{coupling2}. This is demonstrated in Fig.~\ref{fig:2terminal_gen} which displays the resultant conductance in the zero-temperature limit. In the special case $\eta_1^{\rm eff}(\tilde{q})\approx\eta_2^{\rm eff}(\tilde{q})$ and $\tilde{\chi}^{\rm eff}(\tilde{q})\approx\pi/2$ for one of the sectors, the conductance signal is strongly suppressed for that $\tilde{q}$ due to interference effects. Since the value of $\kappa_n$ depends on the overlap between the parafermionic modes with the electronic states in the lead, we assume that $\kappa_1=\kappa_5$ and $\kappa_2=\kappa_4$ as they correspond to conjugate parafermion operators in Eq.~\eqref{coupling2}. In Fig.~\ref{fig:2terminal_gen}(b)-(d) we see that even for non-overlapping parafermions ($\varepsilon=0$) and all $\chi_n=0$, the conductance remains peaked at two finite values of $V_b$. Finally, we note that the optimal peak splitting is no longer given by $\phi=0.19$ due to the peak position displacement by the finite $\tilde{\chi}^{\rm eff}$. 

\section{Supplemental material: Field theory description of the floating SC island}
The techniques adopted in the experiment \cite{kim2020} remove a thin stripe of graphene extending into the bulk of the FQH liquid, thus defining a trench with two counterpropagating gapless modes. NbN is deposited on top of this trench. A low-energy description of this two-parafermion device can be built according to the chiral Luttinger liquid theory for the FQH edge modes by including the induced SC crossed-Andreev pairing between counterpropagating edge modes \cite{stern2012,shtengel2012,vaezi2013,cheng2013,prophecy,burnell2016,kim2017b}. We focus on the Laughlin state at filling $\nu=1/3$. By following a standard bosonization procedure, we introduce two dual gapless bosonic fields satisfying the commutation relation
\begin{equation}
\left[\theta(x),\varphi(y)\right] = -i\frac{\pi}{3} \Theta(x-y)\,, \label{bosonization1}
\end{equation}
where $\Theta(x)$ is the Heaviside step function. The dynamics of the two counterpropagating FQH edge modes beneath the superconductor is determined by the Hamiltonian 
\begin{equation}
H_0=\frac{3v}{2\pi}\int_0^L dx\, \left[(\partial_x\varphi)^2+(\partial_x\theta)^2\right]\,,
\end{equation}
where $v$ is the velocity of the edge modes and $L$ the length of the SC finger. These can be conveniently described in terms of vertex operators. In particular, left and right chiral quasiparticles of charge $e$ are represented by, respectively,
\begin{equation} \label{quasielectron}
\psi_{e,L} = \frac{1}{2\pi}\sqrt{\frac{E_{\rm QH}}{v}}\e^{i3\left(\varphi + \theta\right)} \,,\quad \psi_{e,R} = \frac{1}{2\pi}\sqrt{\frac{E_{\rm QH}}{v}} \e^{i3\left(\varphi - \theta\right)}\,,
\end{equation}
where $E_{\rm QH}$ is the FQH bulk energy gap which plays the role of an ultraviolet cutoff. Typically, $E_{\rm QH}\sim 0.017e^2/\epsilon l_B \sim 1.7$ meV in graphene setups where $\epsilon\approx1$ is the dielectric constant in suspended graphene and $l_B$ is the magnetic length \cite{bolotin2009}.
The crossed-Andreev pairing corresponds to an effective interaction between counterpropagating modes of the form:
\begin{align} \label{pwave}
H_{\Delta}&=-\Delta \int_0^L dx \,  \left[\e^{i\phi_{\rm SC}} \psi_{e,L} \psi_{e,R} + {\rm H.c.}\right] \nonumber\\ 
&=-\frac{\Delta E_{\rm QH}}{2\pi^2 v} \int_0^L dx \, \sin\left(6 \varphi + \phi_{\rm SC}\right)\,,
\end{align}
where $\Delta \propto \Delta_{\rm NbN} \e^{-W/\xi}$ is the modulo of the induced crossed Andreev pairing parameter between $\psi_{e,L}$ and $\psi_{e,R}$. Since these QH fields have the same polarization, their induced pairing crucially relies on the presence of a strong spin-orbit coupling in the SC material, as emphasized in \cite{kim2017,kim2020}. The exponential decay of $\Delta$ with the width $W$ of the SC finger can be deduced from the experimental data on crossed Andreev conversion \cite{kim2017}, and is consistent with theoretical analysis based on hybrid semiconductor-superconductor heterostructures \cite{feinberg2003,danon2015,loss2017}. Here $\xi$ is a length scale matching the coherence length for clean superconductors and depending on the diffusive length in the presence of disorder. The thin SC fingers studied in Ref.~\cite{kim2017} are typically characterized by $\xi \sim 50$ nm and $W$ in the range $50-200$ nm, which is significantly larger than the magnetic length in graphene, $l_B=\sqrt{\hbar/eB}\lesssim 10$ nm for the relevant fields \cite{bolotin2009}. 

We set the phase $\phi_{\rm SC}=\pi/2$ for the sake of simplicity (this phase can be varied by a gauge transformation and does not affect our results). For an isolated SC island, the system is effectively in the gapped FQH state in the regions $x<0$ and $x>L$. This imposes Neumann boundary conditions on the field $\varphi$, corresponding to the absence of a current from the bulk of the FQH state to the edge modes beneath the superconductor. When introducing the contacts with the leads, however, Dirichlet boundary conditions can be dynamically imposed, corresponding to conducting phases.

The Lagrangian of the system expressed in the $\varphi$ field reads
\begin{multline}
\mathcal{L}= \frac{3}{2\pi} \int_{0}^L dx \, \left[ \frac{1}{v} \left(\partial_t \varphi \right)^2 - v \left(\partial_x \varphi \right)^2\right]  \\
+ \frac{\Delta E_{\rm QH}}{2\pi^2 v} \int_0^L dx \, \left[\cos\left(6 \varphi\right) -1\right]\,. \label{actionpf}
\end{multline}
For $\Delta>0$, this corresponds to a sine-Gordon model. In a semiclassical description of the gapped SC phase, the introduction of a fractional quasiparticle in the system roughly corresponds to the creation of a soliton for the sine-Gordon term. Specifically, this soliton describes an excitation interpolating from one classical minimum $\varphi_j = j\pi/3$ of the interaction $H_\Delta$, to the next minimum $\varphi_{j+1}$.
These fractional quasiparticles/solitons have a semiclassical gap $\Delta_{e/3} = \sqrt{8 \Delta E_{\rm QH} / 3 \pi^3}$ (see for example \cite{mussardo}). A quasielectron in the system corresponds instead to an increase of $\pi$ in $\varphi$. Since the fractional solitons are repulsive, a configuration in which $\varphi$ increases by $\pi$ is unstable and it typically decays into three solitons. In our modelling of the blockaded device, however, we neglect this interaction. We approximate in a rough way a charge $e$ excitation as a particle with mass 
\begin{equation}
\Delta_e = 3 \Delta_{e/3} = \sqrt{24\Delta E_{\rm QH} /\pi^3}\,,
\end{equation} 
(thus underestimating its gap) and dispersion $\varepsilon(k) = \sqrt{\Delta_e^2 + v^2k^2}$. Next, we consider finite size effects and open boundary conditions such that we include only a set of discrete states; we consider a momentum discretization $h/(2L)$, such that the corresponding energy discretization is $\delta = v h/(2L)$. For excitation energies beyond the quantum Hall gap, our effective 1D description breaks down and for this reason, within our model, we only consider a set of $E_{\rm QH} /\delta \sim 8 \equiv n_{\rm max}$ excited states. Their corresponding energies are 
\begin{equation}
\varepsilon_n = \sqrt{\Delta_e^2 + \delta^2 \left(n+1/2\right)^2}\,,
\label{eq:qeEn}
\end{equation} with associated annihilation operators $\gamma_n$. We point out, however, that we typically consider ranges of temperature and voltage bias much below $E_{\rm QH}$. For instance, $k_{\rm B} T/ E_{\rm QH}\lesssim 10^{-2}$ for the zero-bias conductance estimates of the blockaded device presented in the main text. Therefore we expect the transport contribution from states beyond the cutoff $E_{\rm QH}$ to be negligible.

In Eq.~\eqref{eq:qeEn} the offset $1/2$ in the quantization index is taken from similar choices in Majorana devices close to the topological phase transition \cite{alicea2016}, and it accounts for confinement effects. Depending on the boundary conditions, however, different offsets could be adopted. In general, the minimum quasielectron energy is $\varepsilon_0 \in \left[ \Delta_e, \sqrt{\Delta_e^2 + \delta^2/4} \right]$.  

The zero-energy parafermionic modes appear in the strong pairing limit, where $\Delta$ flows to strong coupling in the renormalization group (RG) sense. In the ideal case $\Delta E_{\rm QH} \to \infty$, the field $\varphi$ is pinned to one of the six minima of the potential, and may be treated as a discrete operator $\varphi_{\rm SC} = 2\pi j /6$ with $j=0,\ldots,5$ for $0<x<L$.  If the system is isolated and embedded in a FQH state, we can model the external regions for $x<0$ and $x>L$ through strong electron backscattering terms, proportional to $\cos(6\theta)$, that gap the FQH edge modes by pinning the field $\theta$ to six discrete values. The field $\theta$, in the external left and right regions, can thus be considered as two discrete operators $\theta_{L/R}=2\pi m /6$ with $m=0,\ldots,5$ (see a similar description in \cite{barkeshli2014}). In \cite{shtengel2012}, it was shown that the two parafermionic modes can be described, in the low-energy limit, by the vertex operators
\begin{equation} \label{bosonization2}
\alpha_1 = \e^{i(\varphi_{\rm SC} - \theta_L)}\,,\qquad \alpha_2 = \e^{i(\varphi_{\rm SC} - \theta_R)}\,.
\end{equation}
These operators obey the required parafermionic commutation relations, which can be verified by applying the Campbell-Baker-Haussdorf formula and Eq.~\eqref{bosonization1}.
The operators in \eqref{bosonization2} are built via the right chiral combination of the fields $\varphi - \theta$. Analogous and topologically equivalent modes can be built from the left chiral fields. The descriptions of the system in terms of the two chiralities are completely equivalent (they differ only by local operators) and the chirality choice for expressing the parafermions does not imply additional degrees of freedom \cite{prophecy}.

In the case of finite $\Delta$, the coupling $\varepsilon$ of the two parafermions presented in the main text can be derived through an instanton calculation based on the Lagrangian \eqref{actionpf} \cite{burnell2016}. For the parafermions to be well-defined, they must be subgap modes, consistently with $\varepsilon < \Delta_{e/3}$. The parafermion Hamiltonian in Eq.~\eqref{eq:H_pf}, in particular, must be considered a low-energy approximation below the energy scale set by $\Delta_{e/3}$.

The description of the model based on bosonization gives us also the possibility of investigating the RG scaling of the coupling between the metallic leads and the parafermions. For simplicity, let us consider the first term in the tunnelling Hamiltonian \eqref{coupling}, expressed in terms of Majorana modes, for $\chi_1=0$ and in the strong pairing limit $\Delta \to \infty$:
\begin{equation}
H_{c,1} = i \eta_1 \gamma_1 \left(l+l^\dag\right) \propto \eta_1 \sin\left({\Phi_l}-3\varphi_{\rm SC}\right).
\end{equation}
Here $\e^{i\Phi_l} \approx l$ represents the annihilation operator of an electron at the edge of the metallic lead, in such a way that this Hamiltonian describes the hopping of a charge $e$ between the lead and the fractional superconductor. This term is analogous to the coupling between a normal wire and a Majorana mode discussed in Ref.~\cite{fidkowski2012}. As discussed above, in the topological limit of large $\Delta$ the bosonic field $\varphi$ is pinned to the one of the minima $\varphi_{\rm SC}$. Therefore, a simple first order RG analysis reveals that this coupling interaction has scaling dimension $1/2$ \cite{fidkowski2012} and is thus relevant in the RG sense. All the terms in the generalized coupling \eqref{coupling2} are quadratic in the fermionic operators, and behave in the same way. In principle, the tunnelling between the metallic lead and the fractional superconductor could be further generalized by considering interacting terms of higher order in the fermionic degrees of freedom of the lead. This generalization, however, would result in operators with higher scaling dimension, thus negligible, in the RG sense, with respect to Eqs. \eqref{coupling} and \eqref{coupling2}. When relaxing the large-$\Delta$ constraint and considering the effect of the charging energy of the device, the RG analysis becomes more complex, and it results in an RG behaviour of the coupling $H_{c,1}$ which ranges from relevant to marginal (see \cite{kim2017b,oreg2020} for related analyses in the case of fractional quasiparticle transport).

\section{Supplemental material: Estimates of the zero-bias conductance for the floating device}
The conductance of the Coulomb blockaded device is obtained by assuming that the left and right leads are only coupled to the left and right edges of the device, respectively. We consider the electron field $\psi_e= \psi_{e,L} + \psi_{e,R}$ [see Eq.~\eqref{quasielectron}] which annihilates an electron in the paired FQH edge modes beneath the SC finger. We express the coupling between the metallic leads and the floating device as
\begin{equation}
H_c = \sum_{a=L,R} \eta_a l_a^\dag \psi_e(x_a) + {\rm H.c.}\,,
\label{eq:H_c}
\end{equation}
where $a$ labels the left and right leads, $l_a$ is the annihilation operator of an electron at the end of the lead $a$, $x_L\sim 0,\,x_R\sim L$ indicate the edges of the blockaded device, and $\psi_e(x_a)$ annihilates a charge $e$ in the FQH edges in a region localized around $x_a$. The coupling strength at the edge $a$ is $\eta_a$ which, following \cite{vanheck2016}, can be expressed in terms of the dimensionless conductance $g_a$ of the tunnel junction between lead $a$ and the FQH edges (multiplied by $2e^2/h$, $g_a$ yields the contact conductance): 
\begin{equation}
\eta_a=\frac{1}{2\pi}\sqrt{\frac{g_a}{\nu_l\nu_e}}\,.
\end{equation} 
Here $\nu_l$ and $\nu_e$ are the density of states in the leads and in the FQH edge, respectively. In particular, we approximate the latter with $\nu_e \approx \delta^{-1} = 2L/(vh)$. This determines our choice of the tunnelling energies $\eta_a = \frac{1}{2\pi}\sqrt{g_a hv/2\nu_l L}$ in the derivation of the conductances displayed in Fig.~\ref{fig:float} of the main text. By considering $g_L=g_R=0.09$, $v=10^5$ m/s, $\nu_l=1.727$ (meV)$^{-1}$, and $L=1$ $\mu$m, we obtain $\eta_L=\eta_R=0.052$ meV.

In the following, we will base our conductance calculations on rate equations. It is therefore useful to approximate the operator $\psi_e(x)$ in terms of the localized parafermion modes and the quasielectron excitations, by considering the projection of $\psi_e$ on these. The first amplitude is
\begin{multline}
\bra{N_{\rm C},N_e,N',q-3} \psi_e (x_{L/R}) \ket{N_{\rm C},N_e,N',q} \\
=\alpha^3_{1/2} (x_{L/R}) \approx \sqrt{\frac{\nu_e\Delta_e}{2}}\,,
\label{eq:pf_proj}
\end{multline}
where $\alpha_i^3(x)$ is the amplitude at the position $x$ of the normalized zero-energy operator $\alpha^3_i$, which decays exponentially in the bulk approximately with decay length $1/\nu_e\Delta_e$. Concerning the projection of $\psi_e$ on the quasielectron excitations, we approximate the quasielectrons as propagating, non-interacting particles with mass $\Delta_e$ in a p-wave SC system of the kind \eqref{pwave}, described by the Hamiltonian $H_{p} = \sum_ n \varepsilon_n \gamma_n^\dag \gamma_n$, as discussed in the previous section [see Eq.~\eqref{eq:qeEn}]. Based on this, we consider the overlap of $\psi_e$ with the quasielectron excitations $\gamma_n$ of the system to be
\begin{equation}
\psi_e(x_a) = \sum_n u_n^*(x_a) \gamma_n + v_n(x_a) \gamma_n^\dag\,,
\label{eq:psi_e}
\end{equation}
where we are excluding the low-energy localized modes of the form $\alpha^3$, already accounted for. The coefficients $u_n(x_a)$ and $v_n(x_a)$ are derived by a Bogoliubov diagonalization of the quadratic Hamiltonian. By neglecting their position dependence we will adopt the following approximation: $u_n(x_a)v_n^*(x_a) = \Delta_e/2\varepsilon_n$.

Below we outline the methods for deriving the conductances $G_{R}$ and $G_{1e}$ in Eq.~\eqref{eq:Gtot} in the main text. After this, we will also consider the additional conductance $G_{2e}$ caused by the sequential tunnelling of Cooper pairs. This contribution is responsible for the characteristic 2e periodicity of the zero-bias conductance peaks for Majorana devices in their non-topological phases \cite{vanheck2016,albrecht2016}. In the graphene based parafermion devices, however, its contribution is typically negligible, unless we simultaneously consider weak charging energies and short superconducting fingers.

In our summation of the states $\ket{N_{\rm C},N_e,N',q}$ in Eq.~\eqref{eq:Gtot} we apply the following truncations: we consider the sum of states with up to two fractional quasiparticle excitations, $N'=0,1,2$, and up to a single quasielectron excitation, $N_e=0,1$. States with more excitations are indeed thermally suppressed by the low temperature distribution. 

Furthermore, we emphasize that in the estimate of $H_{\rm SC}$ in Eq.~\eqref{eq:H_SC} we considered the gaps $\Delta_e$ and $\Delta_{e/3}$ as the minimum excitation energies for the quasielectron and fractional quasiparticles. This corresponds to neglecting the kinetic energy of these excitations, thus assuming that only the lowest lying energy modes are populated, and considering a system in the thermodynamic limit, such that $\delta\to 0$.

In the following calculations, however, we include also finite-size effects. Therefore, in the estimates of the thermal weights in \eqref{eq:Gtot}, we minimize the quasielectron and fractional quasiparticle energies by the minima $\varepsilon_0$ and $\varepsilon_0'\equiv \sqrt{\Delta_{e/3}^2+\delta^2/4}$, respectively. 

Concerning the kinetic energy of the excitations, we emphasize that we neglect it only in the estimate of the thermal weights, whereas in the calculations of the tunnelling rates below, the dispersion \eqref{eq:qeEn} has been fully taken into account. 

Finally, given the truncation $N_e=0,1$, it is convenient to introduce the excess electron number $N=2N_{\rm C}+N_e$, as customarily done in the study of Majorana devices \cite{vanheck2016}. In this respect, even values of $N$ label states without quasielectron excitations, whereas odd values correspond to the presence of a single quasielectron excitation. Based on these approximations, we hereafter simply refer to the states $\ket{N,N',q}$ and we rewrite $H_{\rm SC}$ as
\begin{multline}
H_{\rm SC}(N,N',q,n_g) = E_C \left(N + N'/3+q/3-n_g\right)^2 \\ + H_{\rm pf}(q) +  \varepsilon_0\left[1-(-1)^N\right]/2 + \varepsilon_0' N' .
\label{eq:H_SC2}
\end{multline}

The results displayed in the main text are obtained by summing the states with $N\in \left[-6,4\right]$ and $N'=0,1,2$ in Eq.~\eqref{eq:Gtot}.

\subsection{Resonant tunnelling}
\begin{figure}[t]
\centering
\includegraphics[width=\linewidth]{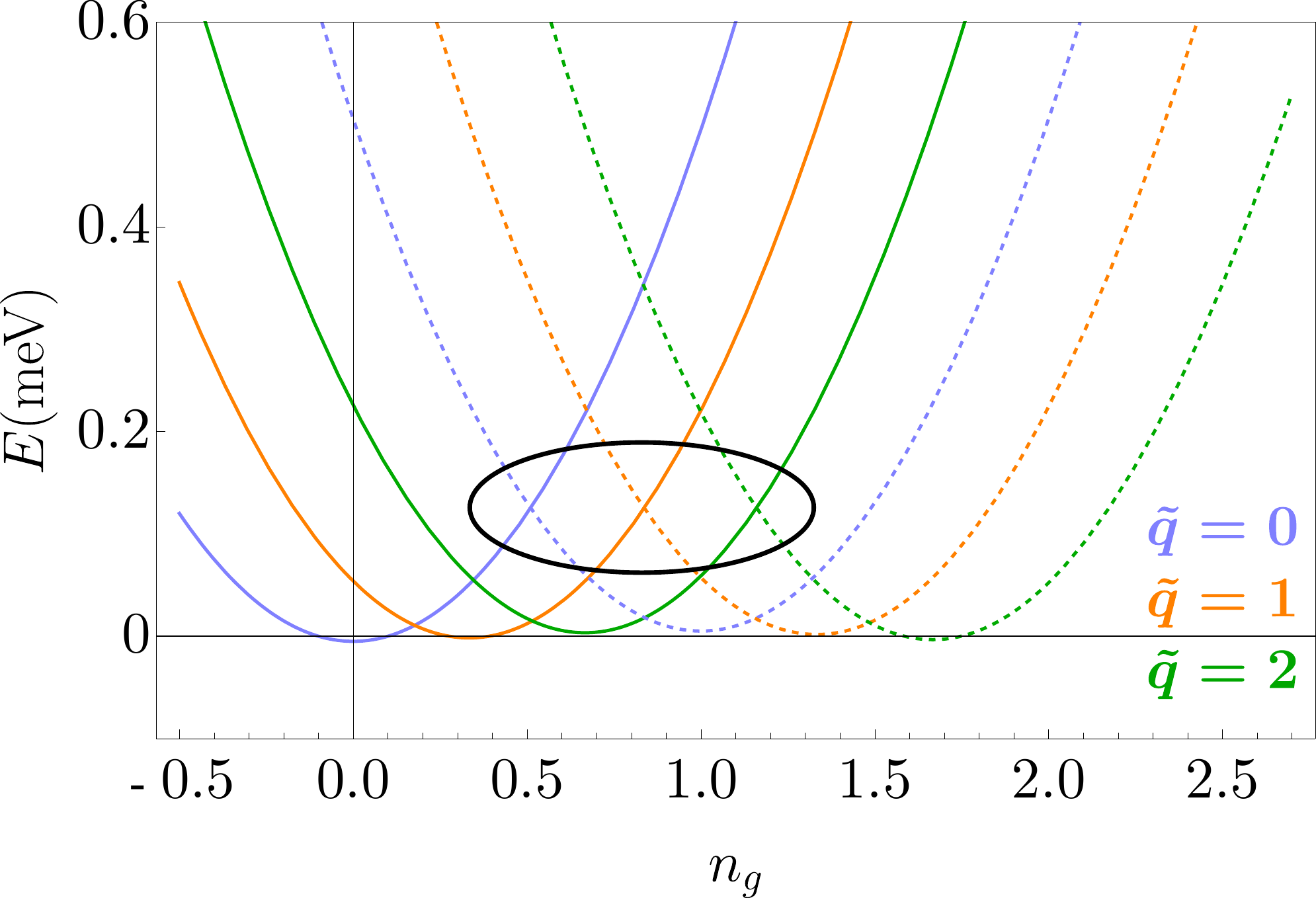}
\caption{Eigenenergies of $H_{\rm SC}$ as a function of the gate parameter $n_g$ for $q=0,1,2$ (full) and $q=3,4,5$ (dashed). For simplicity we plot only the parabolas with $N'=0$; only the "even parabolas" $N=0$ without quasielectron excitations are visible in this low-energy range. Parameter values are the same as in Fig.~\ref{fig:float}(c) with $\Delta=0.3$ meV.}
\label{fig:parab_R}
\end{figure}

When the voltage bias between the leads is small, resonant electron tunnelling requires the two involved states, $\ket{N,N',q}$ and $\ket{N,N',q\pm3}$, to be close in energy. To visualize this, we plot in Fig.~\ref{fig:parab_R} energy parabolas of Eq.~\eqref{eq:H_SC2} for six different values of $q$. The three crossings of $q$ and $q+3$ inside the black circle set the values of $n_g$ for which zero-bias conductance is possible. 

Our aim is to estimate the resonant tunnelling transport in the regime in which these pairs of states are isolated, in energy, from the excited states with odd $N$ appearing at energies greater than $\Delta_e$. In particular, we demand that $\varepsilon \ll E_C \ll \Delta_e$. Under this assumption, we will again use the Weidenm\"uller formula for the calculation of $G_{R}$, but now with an explicitly charge conserving coupling (thus neglecting Andreev processes). Inserting the projection in Eq.~\eqref{eq:pf_proj} into the coupling Hamiltonian in Eq.~\eqref{eq:H_c}, we can write it as $H_c=\sum_{a=L,R}c^\dag W'\underline{f}+{\rm H.c.}$ Here $c$ is the fermionic operator $c=(\gamma_1+i\gamma_2)/2$ defined previously and $f=(l_L,l_R)$ is the vector of annihilation operators associated with electrons at the left and right tunnel contact. The coupling matrix $W'$ is
\begin{equation}
W'= \sqrt{\frac{\nu_e\Delta_e}{2}}\begin{pmatrix}
\eta_L & \eta_R
\end{pmatrix}\,.
\end{equation} 
The factor in front of the matrix indeed comes from the localization of the parafermions. We represent these by considering only the two states $\ket{N,N',q},\ket{N,N',q-3}$, modelled as a two-level system $H_{\rm pf}'=\Delta_q(c^\dag c-1/2)$ with the energy splitting $\Delta_q=2E_C(N+(N'+q)/3-1/2-n_g)-4\varepsilon\cos(\pi q/3+\phi)$. By applying the Weidenm\"uller formula \eqref{weiden}, we find the scattering matrix. The corresponding differential conductance at zero voltage bias reads
\begin{equation}
\tilde{G}_{R} = \frac{e^2}{h}\frac{g_L g_R}{4(2\pi)^2}\frac{\Delta_e^2}{\Delta_q^2 + (g_L+g_R)^2\Delta_e^2/(8\pi)^2}\,.
\label{eq:G_R}
\end{equation}
The expression is valid when the level broadening induced by the contact to the leads is much greater than the temperature: $T\ll(g_L+g_R) \Delta_e /8\pi$ \cite{vanheck2016} and when $\Delta_e \gg E_C$ such that states with $N'\neq0$ and odd $N$ are well separated in energy from the lowest-energy states $N'=0$ and $N$ even. Note that for symmetric contacts ($g_L=g_R$), the peak conductance  in Eq.~\eqref{eq:G_R} is given by $\tilde{G}_{R}=e^2/h$, in accordance with the teleportation peak conductance derived by Fu for the MZM case \cite{fu2010}. This quantization, however, is lost when including the Boltzmann weights as in Eq.~\eqref{eq:Gtot}, and the magnitude of $G$ becomes very sensitive to temperature. We recognize that by including only the states $\ket{N,N',q},\ket{N,N',q-3}$ in the calculation above (and not $\ket{N,N',q+3}$) we avoid double-counting when summing over $N,q$.

\subsection{Sequential tunnelling of single electrons}
\begin{figure}
\centering
\includegraphics[width=\linewidth]{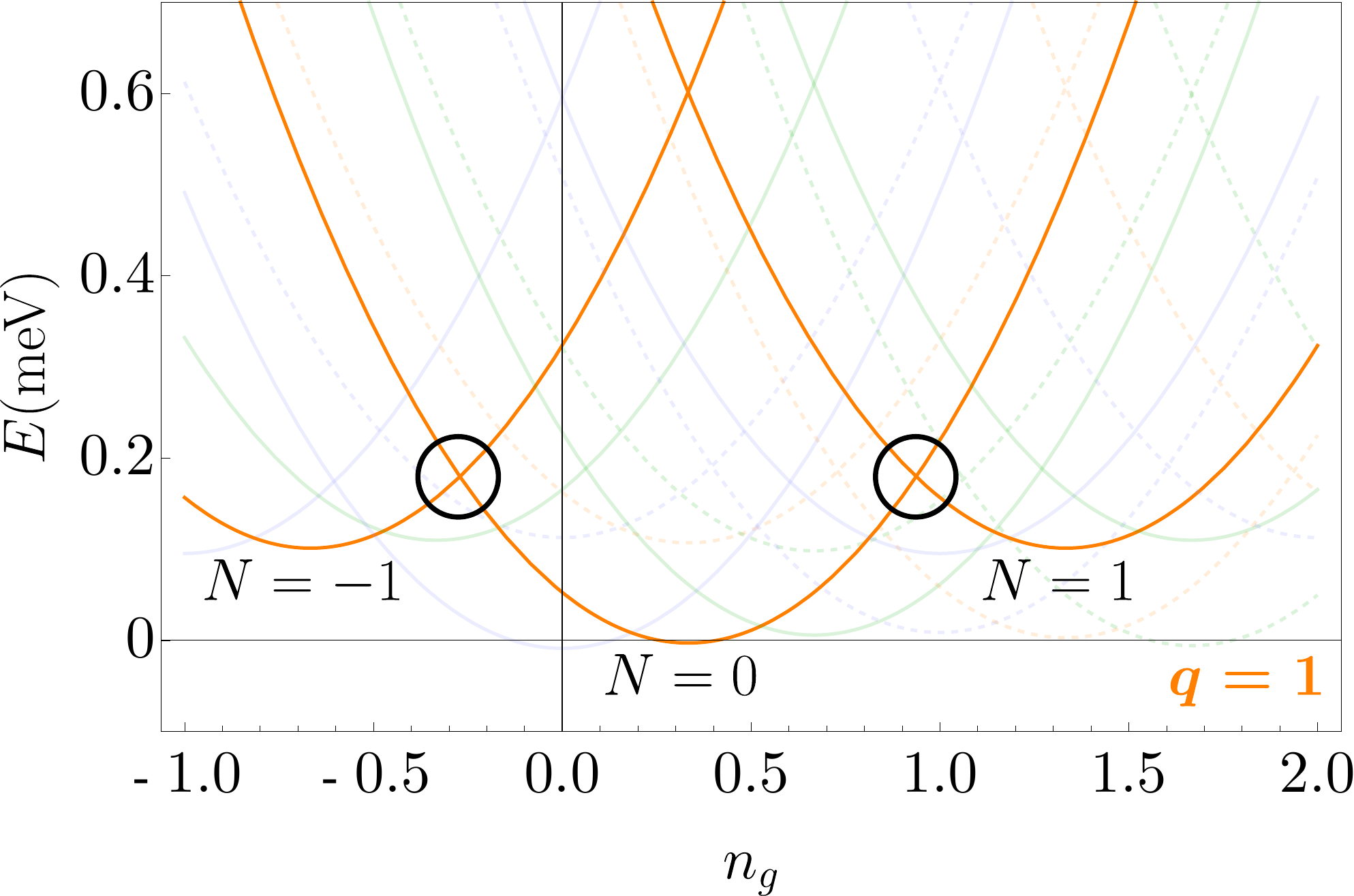}
\caption{Eigenenergies of $H_{\rm SC}$ as a function of gate parameter $n_g$. In the opaque background are states with $q\neq1$ such that the two $1e$ crossings marked by black circles are better visible. As in Fig.~\ref{fig:parab_R} $N'=0$. Parameter values are the same as in Fig.~\ref{fig:float}(b) with $\Delta=10^{-4}$ meV.}
\label{fig:parab_1e}
\end{figure}

Sequential tunnelling is relevant in the regime $\Delta_e < E_C$. In Fig.~\ref{fig:parab_1e}, we plot the energies of Eq.~\eqref{eq:H_SC2} for $N=-1,0,1$, $N'=0$, and $q=0,\ldots,5$. The states $\ket{N,N'=0,q=1}$ are displayed in bright colours to exemplify the crossings of states $\ket{N,N',q}$ and $\ket{N\pm1,N',q}$ that mark values of $n_g$ where zero-bias conductance peaks are expected. 

First, we consider the crossing between $\ket{N}$ and $\ket{N+1}$ with $N$ even (an example of this is the right circle in Fig.~\ref{fig:parab_1e}). As a lowest-order approximation of the rate for an electron in lead $a$ with momentum $p$ to scatter into the system of $N$ electrons and excite a quasielectron with energy $\varepsilon_n$, we use Fermi's golden rule:
\begin{equation}
\Gamma^{a,p}_{N\to N+1,n}=\frac{2\pi}{\hbar}\left|\prescript{}{a}{\bra{0}}\prescript{}{n}{\bra{1}}\, H_c\, \ket{0}_n\ket{p}_a\right|^2 \delta(\mathcal{E}_{N+1}-\xi_p+\varepsilon_n)\,.
\end{equation}
Here $\xi_p$ is the energy of the lead electron and
\begin{equation} \label{EN1}
\mathcal{E}_{N+1}=2E_C(N+(N'+q)/3+1/2-n_g)
\end{equation}
is the difference in charging energy between the states with $N+1$ and $N$ electrons. $\ket{1}_n$ labels the state with $N+1$ electron charges whereof one is the excited quasielectron.
We employ Eq.~\eqref{eq:psi_e}, the momentum representation of $l_a$ (with wave function $\phi_p(x_a)$), and follow the normalization in \cite{vanheck2016} ($\eta_a^2|u_n(x_a)|^2|\phi_p(x_a)|^2=g_a \delta \delta_a/2(2\pi)^2$, where $\delta$ is the quantized energy defined in the previous section and $\delta_a$ is the level spacing in lead $a$). This way, the rate becomes
\begin{equation}
\Gamma^a_{N\to N+1,n}=\frac{g_a\delta}{4\pi\hbar} n_{\rm F}(\mathcal{E}_{N+1}+\varepsilon_n-eV_a)\,.
\end{equation}
Here $n_{\rm F}$ the Fermi distribution function and $V_a$ is the voltage applied to lead $a$. Similarly, the rate for the opposite process is: $\Gamma^a_{N+1,n\to N}=\frac{g_a\delta}{4\pi\hbar} (1-n_{\rm F}(\mathcal{E}_{N+1}+\varepsilon_n-eV_a))$. 

Analogously, we can consider the transitions between $\ket{N}$ and $\ket{N+1}$ with $N$ odd (left circle in Fig.~\ref{fig:parab_1e}). In this case, an electron from lead $a$ with momentum $p$ can bring the system from the state with $N$ electron charges, one of them being an excited quasielectron, into a state without the excitation and $N+1$ electron charges. The rate for this is: $\Gamma_{N,n\to N+1}^a=\frac{g_a\delta}{4\pi \hbar}n_{\rm F}(\mathcal{E}_{N+1}-\varepsilon_n-eV_a)$, and $\Gamma_{N+1\to N,n}^a$ is defined correspondingly. We find an expression for the current across the SC island due to sequential tunnelling of single electrons by using a steady state solution to a simplified set of rate equations. This implies that when calculating the current close to the crossing of $\ket{N}$ and $\ket{N+1}$, the occupation probabilities of all other states are set to zero \cite{vanheck2016}. From this, we obtain the zero-bias differential conductance
\begin{equation}
\mathcal{G}_{1e}^{\rm o/e}=\frac{e^2}{2h}\frac{g_L g_R \frac{\delta}{k_{\rm B}T}}{g_L+g_R}\frac{\sum_n\left(1+\e^{\left[\varepsilon_n\mp \mathcal{E}_{N+1}\right]/k_{\rm B}T}\right)^{-1}}{1+\sum_n \e^{-\left[\varepsilon_n\mp \mathcal{E}_{N+1}\right]/k_{\rm B}T}}\,.
\end{equation}
Here the $\rm o/e$ superscript refers to whether $N$ is odd or even. The energies $\varepsilon_n$ and $\mathcal{E}_{N+1}$ are defined in Eqs.~\eqref{eq:qeEn} and \eqref{EN1}, respectively. The conductance appearing in the weighted sum in Eq.~\eqref{eq:Gtot} is
\begin{equation}
\tilde{G}_{1e}(n_g-(N+(N'+q)/3))= \mathcal{G}_{1e}(N-1)+\mathcal{G}_{1e}(N)\,.
\end{equation}
When $T$ is the smallest energy scale of the system, the sixfold pattern characterizing the sequential tunnelling zero-bias peaks reveals the structure of the energy level crossings displayed in Fig. \ref{fig:parab_1e}.

\subsection{Sequential tunnelling of Cooper pairs}
\begin{figure}[t]
\centering
\includegraphics[width=\linewidth]{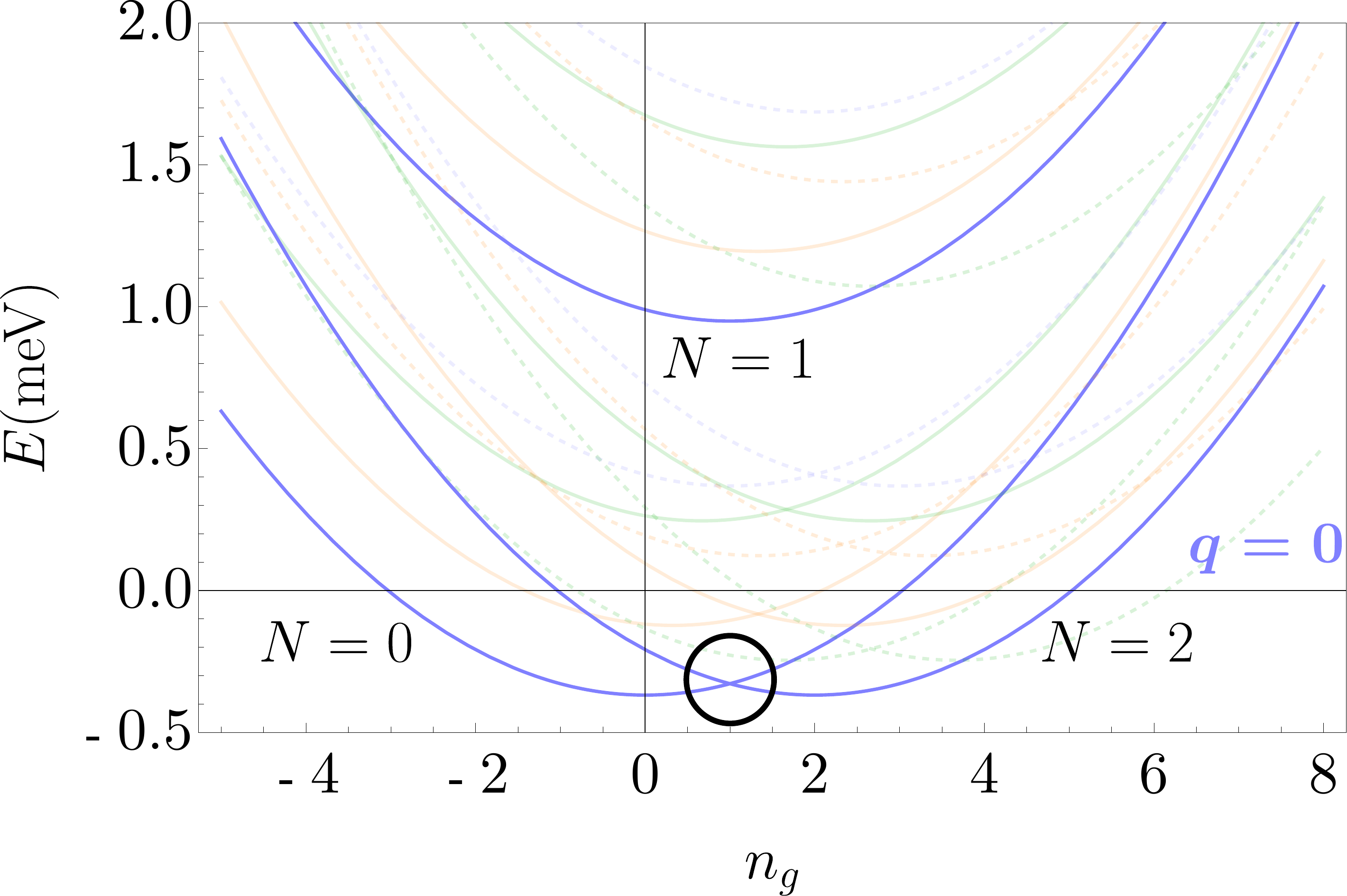}
\caption{Eigenenergies of $H_{\rm SC}$ as a function of gate parameter $n_g$. In the opaque background are states with $q\neq0$ ($N=0,1,2$) in order to highlight the $2e$ crossing marked by a black circle. As in Fig.~\ref{fig:parab_R} $N'=0$. Parameter values are: $\Delta=0.3$ meV, $E_{\rm QH}=1.72$ meV, $v=10^5$ m/s, $T=0.15$ K, $L=0.1$ $\mu$m, and $E_C=0.04$ meV.}
\label{fig:parab_2e}
\end{figure}

\begin{figure}[t]
\centering
\raisebox{0.1ex}{\includegraphics[width=0.479\columnwidth]{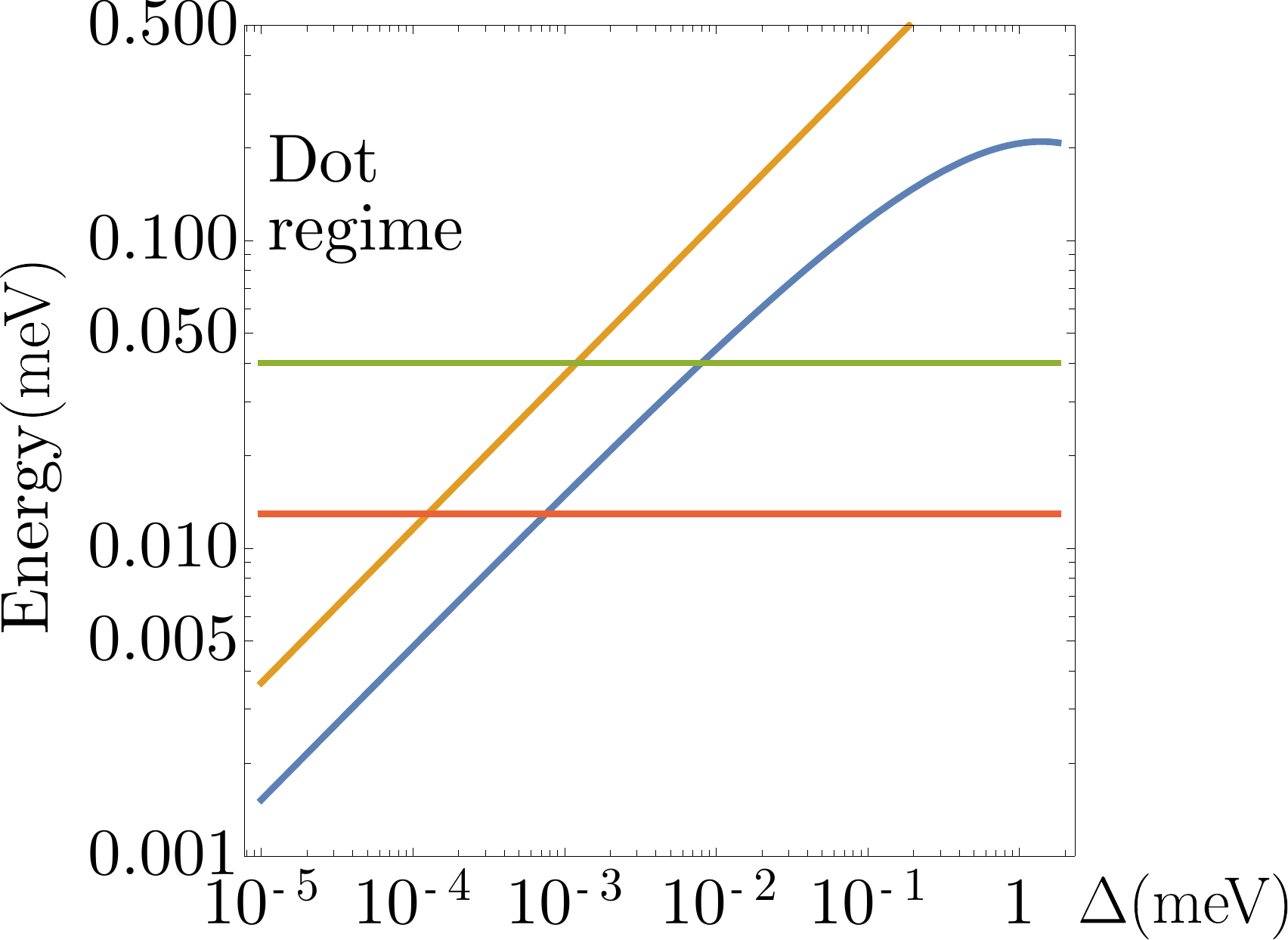}}
\llap{\parbox[b]{8.45cm}{\color{black}(a)\\\rule{0ex}{2.8cm}}}
\hfill
\includegraphics[width=0.48\columnwidth]{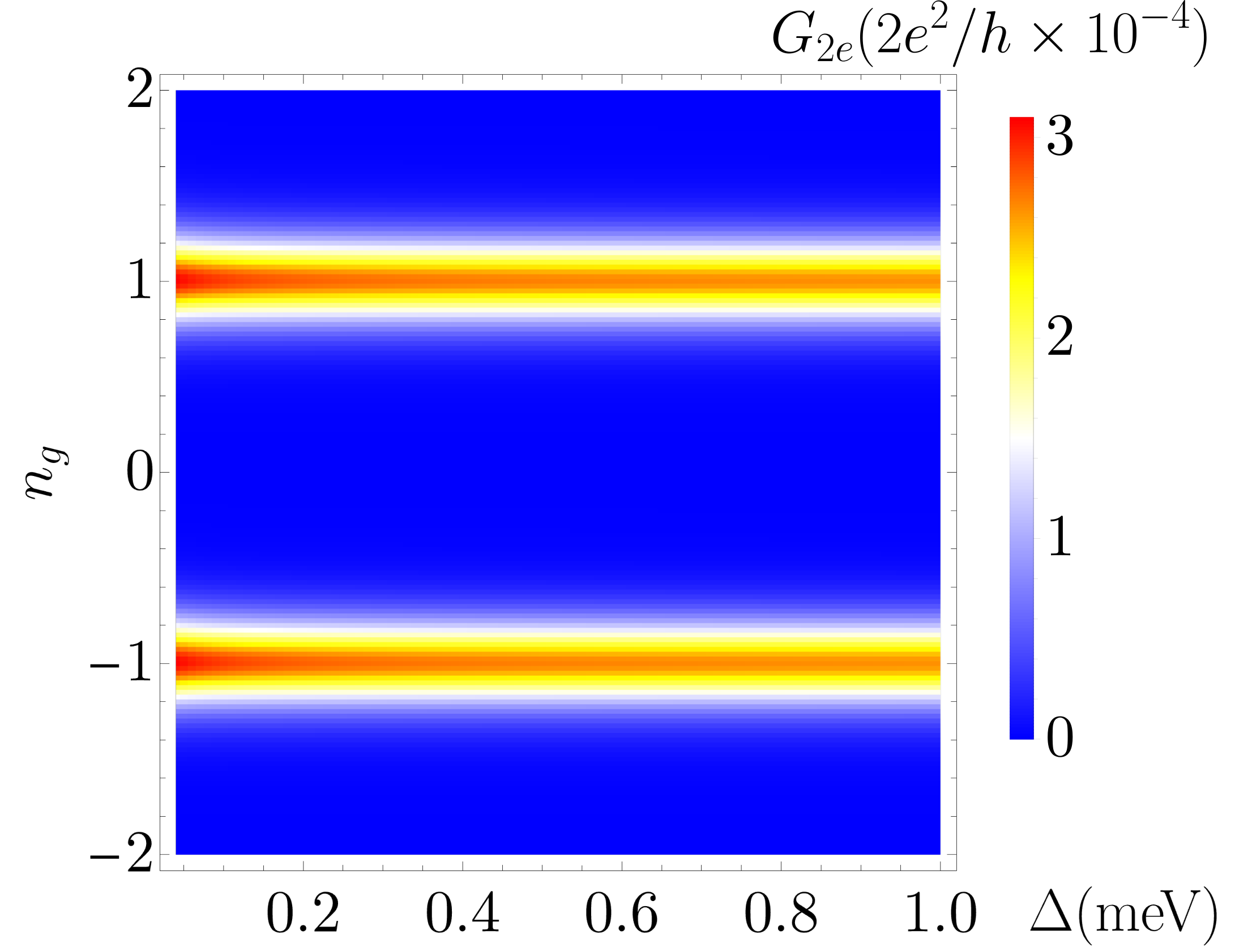}
\llap{\parbox[b]{8.2cm}{\color{black}(b)\\\rule{0ex}{2.8cm}}}
\hspace*{0.5ex}
\caption{Blockaded device. (a) Comparison (in logarithmic scale) between energy scales $\varepsilon$ (blue), $\Delta_e$ (orange), $T=0.15$ K (red), and $E_C$ (green) for same parameter values as in Fig.~\ref{fig:float}(a), but with $L=0.1$ $\mu$m and $E_C=0.04$ meV, i.e. a shorter SC island with smaller charging energy. (b) Zero-bias conductance $G_{2e}$ due to the Cooper pair sequential tunnelling for parameters used in (a).}
\label{fig:cond2e}
\end{figure}

Here we consider the transport contribution from Andreev sequential tunnelling of Cooper pairs, which is relevant for weak charging energy $E_C \ll \varepsilon$, see Fig.~\ref{fig:cond2e}(a). In Fig.~\ref{fig:parab_2e} we again draw energy parabolas, now for $\ket{N,N'=0,q=0}$ with $N=0,1,2$.

In order to find $\Gamma_{N\to N+2}$ and $\Gamma_{N+2\to N}$, the rates for adding or subtracting a Cooper pair to the quantum Hall edges, respectively, we need the amplitudes for the Andreev reflection processes. These processes yield zero-bias conductance peaks at values of $n_g$ close to odd integers.

The rates are estimated by neglecting the population of states with unpaired quasielectron excitations and thus we consider only $N$ to be even. To second order in the tunnelling Hamiltonian, the amplitude for absorbing two electrons with momenta $p_1,p_2$ into the edge modes from lead $a$ is
\begin{multline}
A^{a,p_1,p_2}_{N\to N+2} =\eta_a^2\phi_{p_1}(x_a)\phi_{p_2}(x_a)\sum_n u_n(x_a) v_n^*(x_a) \\
\times \left(\frac{1}{-\mathcal{E}_{N+1}+\xi_{p_1}-\varepsilon_n}+\frac{1}{-\mathcal{E}_{N+1}+\xi_{p_2}-\varepsilon_n}\right)\,,
\end{multline}
and similarly for $A^{a,p_1,p_2}_{N+2\to N}$. In analogy with the Majorana setups \cite{vanheck2016}, we consider a low-temperature limit and approximate $\xi_{p_1},\xi_{p_2}\approx0$, such that the above expression simplifies into
\begin{equation}
A^{a,p_1,p_2}_{N\to N+2} = \eta_a^2\phi_{p_1}(x_a)\phi_{p_2}(x_a)\sum_n\frac{\Delta_e}{2\varepsilon_n}\frac{-2}{\mathcal{E}_{N+1}+\varepsilon_n}\,.
\end{equation} 
In the regime of interest for the Cooper pair tunnelling, the charging energy difference of the involved states is smaller than the quasiparticle gap: $-\mathcal{E}_{N+1} < \Delta_e$. Based on this assumption and taking the continuum limit $n_{\rm max}\to\infty$ [see above Eq.~\eqref{eq:qeEn}], we obtain
\begin{equation}
|A^{a,p_1,p_2}_{N\to N+2}|^2\approx \frac{g_a^2\delta_a^2}{(2\pi)^4}\frac{4\Delta_e^2}{\Delta_e^2-\mathcal{E}_{N+1}^2}\arctan^2\left(\sqrt{\frac{\Delta_e-\mathcal{E}_{N+1}}{\Delta_e+\mathcal{E}_{N+1}}}\right)\,.
\end{equation}
With this, we can find the scattering rates
\begin{multline}
\Gamma^a_{N\to N+2}=\frac{2\pi}{\hbar}|A^{a}_{N\to N+2}|^2\sum_{p_1,p_2}\delta(\xi_{p_1}+\xi_{p_2}-\mathcal{E}_{N+2}) \\
\times n_{\rm F}(\xi_{p_1}-eV_a)n_{\rm F}(\xi_{p_2}-eV_a)\,,
\end{multline}
and similarly for $\Gamma^{a}_{N+2\to N}$ with $n_{\rm F}\to1-n_{\rm F}$. $\mathcal{E}_{N+2} = 4E_C(N+(N'+q)/3+1-n_g)$ is the difference in charging energy between states with $N$ and $N+2$ electrons. We use a result from \cite{vanheck2016} for the $2e$ current contribution and find the zero-bias differential conductance to be
\begin{multline}
\mathcal{G}_{2e} = \frac{e^2}{2h} \frac{g_L^2 g_R^2}{g_L^2+g_R^2} \frac{1}{\pi^2} 
\frac{\Delta_e^2}{\Delta_e^2-\mathcal{E}_{N+1}^2}\\
\times
\arctan^2\left(\sqrt{\frac{\Delta_e-\mathcal{E}_{N+1}}{\Delta_e+\mathcal{E}_{N+1}}}\right)\frac{\mathcal{E}_{N+2}/k_{\rm B}T}{\sinh\left[\mathcal{E}_{N+2}/k_{\rm B}T\right]}\,.
\end{multline}
Including weak quasiparticle poisoning in the same way as for the resonant and $1e$ cases [see Eq.~\eqref{eq:Gtot}] the conductance contribution from Cooper pair tunnelling is
\begin{equation}
G_{2e}= \sum_{N,N',q}\frac{\e^{-H_{\rm SC}(N,N',q,n_g)/T}}{Z} \left(\mathcal{G}_{2e}(N)+\mathcal{G}_{2e}(N-2)\right)\,.
\end{equation}
In Fig.~\ref{fig:cond2e}(b) we plot $G_{2e}$ for a short superconducting finger ($L=0.1$ $\mu$m) and a weak charging energy, i.e. $\Delta_{e}>\varepsilon \gg E_C \gg T$ where we expect the Cooper pair tunnelling channel to be relevant. The pattern shows a single peak repeating with $2e$ periodicity as opposed to the sixfold regime in the resonant channel [Fig.~\ref{fig:float}(c)]. The considered SC finger is ten times shorter than that of the main text. However, for thin superconductors we expect the typical charging energies to be larger than $\varepsilon$ for any $\Delta$, such that this Andreev dominated regime may be hard to observe and the sixfold pattern dominates for all intermediate pairings [as in Fig.~\ref{fig:float}(e) for $\Delta \lesssim 0.3$ meV].

\end{document}